%% file: HIN-25-008_temp.tex
\pdfoutput=1
\documentclass[11pt,twoside,a4paper,cmspaper,final,collab]{cms-tdr}
\def\svnVersion{b9668b1}\def\svnDate{2026/03/18}\def\cmsCernNoTag{CERN-EP-2025-226}\def\cmsCernDate{\today}\def\cmsMessage{Published in Physical Review Letters as \href{http://dx.doi.org/10.1103/89sf-9t1x}{\doi{10.1103/89sf-9t1x}.}}
\begin{document}\cmsNoteHeader{HIN-25-008}

\newcommand{\pp}{\ensuremath{\Pp\Pp}\xspace}
\newcommand{\pPb}{\ensuremath{\Pp\text{Pb}}\xspace}
\newcommand{\PbPb}{\ensuremath{\text{PbPb}}\xspace}
\newcommand{\XeXe}{\ensuremath{\text{XeXe}}\xspace}
\newcommand{\OO}{\ensuremath{\text{OO}}\xspace}
\newcommand{\raa}{\ensuremath{R_{\mathrm{AA}}}\xspace}
\newcommand{\dSigdPtAA}{\ensuremath{\frac{\rd \sigma^{\mathrm{AA}}}{\rd \pt}}\xspace}
\newcommand{\dSigdPtpp}{\ensuremath{\frac{\rd \sigma^{\pp}}{\rd \pt}}\xspace}
\newcommand{\dNdEta}{\ensuremath{\rd N_{\text{ch}}/\rd\eta}\xspace}

\newlength\cmsFigWidth
\ifthenelse{\boolean{cms@external}}%
{   
\renewcommand{\appendixname}{End Matter}
\newcommand{\addmatter}[1]{\appendixname~\ref{#1}}
\newcommand{\suppmat}[1]{the supplemental material~\cite{suppMatBib}}
\providecommand{\cmsLeft}{upper\xspace}
\providecommand{\cmsRight}{lower\xspace}
\setlength{\cmsFigWidth}{0.5\textwidth}
}
{%
\newcommand{\addmatter}[1]{\appendixname~\ref{#1}}
\newcommand{\suppmat}[1]{\appendixname~\ref{#1}\nocite{suppMatBib}}
\providecommand{\cmsLeft}{left\xspace}
\providecommand{\cmsRight}{right\xspace}
\setlength{\cmsFigWidth}{0.75\textwidth}
}

\cmsNoteHeader{HIN-25-008}
\title{Observation of suppressed charged-particle production in ultrarelativistic oxygen-oxygen collisions}

\date{\today}
\abstract{
A hot and dense state of nuclear matter, known as the quark-gluon plasma, is created in collisions of ultrarelativistic heavy nuclei.  Highly energetic quarks and gluons, collectively referred to as partons, lose energy as they travel through this matter, leading to suppressed production of particles with large transverse momenta (\pt).  Conversely, high-\pt particle suppression has not been seen in proton-lead collisions, raising questions regarding the minimum system size required to observe parton energy loss.  Oxygen-oxygen (OO) collisions examine a region of effective system size that lies between these two extreme cases.  The CMS detector at the CERN LHC has been used to quantify charged-particle production in inclusive \OO collisions for the first time via measurements of the nuclear modification factor (\raa).  The \raa is derived by comparing particle production to expectations based on proton-proton (\pp) data and has a value of unity in the absence of nuclear effects. The data for \OO and \pp collisions at a nucleon-nucleon center-of-mass energy $\sqrtsNN=5.36\TeV$ correspond to integrated luminosities of 6.1\nbinv and 1.02\pbinv, respectively.  The \raa is below unity with a minimum of $0.69\pm0.04$ around $\pt=6\GeV$.  The data exhibit better agreement with theoretical models incorporating parton energy loss as compared to baseline models without energy loss.
}

\hypersetup{
pdfauthor={CMS Collaboration},
pdftitle={Observation of suppressed charged-particle production in ultrarelativistic oxygen-oxygen collisions},
pdfsubject={CMS},
pdfkeywords={CMS, heavy ions, jet quenching, nuclear modification factor, quark-gluon plasma}}

\maketitle 

In high-energy collisions of heavy nuclei, a medium consisting of a hot and dense state of nuclear matter with deconfined quarks and gluons, known as the quark-gluon plasma (QGP), is created~\cite{Busza:2018rrf,Shuryak:1980tp}.  The QGP exhibits many striking properties, including fluid-like collective behavior as well as mass- and flavor-dependent modification of particle production when compared to expectations based on proton-proton (\pp) data~\cite{CMS:2024krd,ALICE:2022wpn,BRAHMS:2004adc,PHOBOS:2004zne,STAR:2005gfr,PHENIX:2004vcz}.  Understanding the dynamics of these phenomena and how they emerge from the underlying theory of the strong force, quantum chromodynamics (QCD)~\cite{Gross:1973id,Politzer:1973fx}, is a key goal of modern high-energy nuclear physics.

The microscopic structure of the QGP at short length scales, where individual quark and gluon degrees of freedom are most salient, is experimentally accessible through the examination of high-momentum scattering processes that can happen concurrently with QGP production~\cite{Gyulassy:1990ye}.  The scattered quarks and gluons, collectively known as partons, can interact with the QGP and lose energy via collisional and radiative processes~\cite{Cao:2020wlm}.  This parton energy loss causes ``jet quenching", a term used to describe the resulting modification of yields, energies, and spatial distributions of particles originating from the initial scattered parton~\cite{Connors:2017ptx}.  

In hadron collider experiments, particles having transverse momentum (\pt) greater than a few \GeV are almost entirely produced via the fragmentation and hadronization of high-energy partons.  Thus, high-\pt particle spectra are sensitive to parton energy loss effects.  Modifications of these spectra are typically quantified using a ratio between the spectra of interest and a baseline found using \pp collisions at the same center-of-mass energy per nucleon pair (\sqrtsNN).  Specifically, the relevant \pp spectrum is scaled to account for the larger number of nucleon-nucleon interactions in a nucleus-nucleus collision.  Early measurements of this ratio, known as the nuclear modification factor (\raa), at the CERN Super Proton Synchrotron~\cite{WA98:2001lnw,dEnterria:2004cly}, the Relativistic Heavy Ion Collider (RHIC)~\cite{BRAHMS:2004adc,PHOBOS:2004zne,STAR:2005gfr,PHENIX:2004vcz}, and the CERN LHC~\cite{CMS:2012aa,ALICE:2012aqc,ATLAS:2015qmb} were among the first indications of jet quenching at their respective collision energy scales.  More recent charged-particle \raa measurements at the LHC using lead ($^{208}$Pb) collisions at $\sqrtsNN=5.02\TeV$~\cite{CMS:2016xef,ALICE:2018vuu,ATLAS:2022kqu} and xenon ($^{129}$Xe) collisions at 5.44\TeV~\cite{CMS:2018yyx,ALICE:2018hza,ATLAS:2022kqu} found \raa to be in the range of 0.2--0.3 for particles with \pt around 8\GeV, indicating significant jet quenching effects in collisions of heavy nuclei.

In small collision systems, \eg, \pp and \pPb, observations of QGP-like effects including collectivity~\cite{CMS:2012qk} and the enhanced production of strange hadrons~\cite{CMS:2019isl,ALICE:2016fzo}, have prompted questions of whether a small droplet of QGP can be produced in such collisions~\cite{Nagle:2018nvi}.  However, jet quenching has not been observed in these systems~\cite{CMS:2025jbv,CMS:2015ved,ATLAS:2022iyq,ALICE:2017svf}, and the \pPb \raa is close to unity at particle $\pt>2\GeV$~\cite{CMS:2016xef,ALICE:2014nqx,ATLAS:2022kqu}. A 20\% suppression of the $\PGpz$ \raa in deuteron-gold collisions was seen at RHIC when selecting events producing large particle multiplicities~\cite{PHENIX:2023dxl}, but questions remain regarding the degree to which these event selections may cause measurement biases~\cite{Perepelitsa:2024eik}. In all of these systems, the absence of a clear jet quenching signal could be attributable to a lack of QGP production, or alternatively, the very short path length a parton would have to traverse to exit a QGP droplet.  It has been proposed to experimentally test the latter hypothesis using collisions of light nuclei having low nuclear mass number ($\text{A}$).  Such collisions will potentially create a QGP volume in between those of \pPb and heavier ion collisions~\cite{Huss:2020dwe}.  These intermediate systems are also less sensitive to initial-state energy density fluctuations that hinder theoretical descriptions of smaller systems~\cite{Brewer:2021kiv,Schenke:2020mbo}.  Theoretical models incorporating jet quenching have predicted charged-particle \raa suppression that should be experimentally accessible, of the order of 25\% for $\text{A}=16$ at LHC energies~\cite{Huss:2020whe,Huss:2020dwe,Zakharov:2021uza,Ke:2022gkq}. 

In 2025, the LHC delivered $\TeV$-scale collisions of $^{16}$O nuclei for the first time.  This Letter presents the first measurement of the \pt-differential cross section of charged-particle production in oxygen-oxygen (\OO) collisions at $\sqrtsNN = 5.36\TeV$ using data collected by the CMS experiment.  A reference spectrum measured using \pp data from 2024 at the same collision energy is also reported.  The charged-particle \raa for inclusive \OO collisions is then constructed using the following definition:
\begin{equation}
\raa = \frac{1}{\text{A}^2}\left[\left(\dSigdPtAA\right)\bigg{/} \left(\dSigdPtpp\right)\right],
\label{eq1}
\end{equation}
where $\sigma^{\mathrm{AA}}$ ($\sigma^{\pp}$) is the cross section for charged-particle production in nucleus-nucleus (\pp) collisions.  Notably, because the prefactor accounting for the number of nucleons in $^{16}$O is exact, this formulation does not contain any model-dependent uncertainties, \eg, from Glauber modeling~\cite{dEnterria:2020dwq,Loizides:2016djv}.  The \OO data are compared to other collision systems at similar \sqrtsNN, as well as predictions from theoretical models.  Tabulated results are provided in the HEPData record for this analysis~\cite{hepdata}.

A complete description of the CMS apparatus is reported in Refs.~\cite{CMS:2008xjf,CMS:2023gfb}.  A description of the Monte Carlo (MC) simulations used to evaluate detector performance can be found in \addmatter{endmatter:suppMat1}.  

The \pp data were collected using a trigger that only required two coincident colliding bunches in the detector, while the \OO data were collected using a trigger that searched for one or more energy deposits of at least 7\GeV in either of the forward hadron (HF) calorimeters. The total integrated luminosity utilized was $1.02\pm0.03\pbinv$ for the \pp sample and $6.1\pm0.2\nbinv$ for the \OO sample. These values are determined by combining two independent measurements. The first is a preliminary analysis of van der Meer scan~\cite{vanderMeer:296752,CMS:2025rzq} data.  A second luminosity determination was conducted using a method that compares observed $\PZ$ boson yields to a well-calibrated reference dataset~\cite{CMS:2023pad}. The luminosities extracted from both methods are in good agreement with each other.

The average number of interactions per bunch crossing ($\mu$), typically referred to as pileup, was 4.7 and 0.14 for the \pp and \OO datasets, respectively.  While pileup effects are negligible for \OO collisions, the \pp data are corrected for pileup effects by repeating the measurement using subsets taken at different times.  These subsets sample $\mu$ ranging from 1.0 to 7.6 and are used to extrapolate to $\mu\approx0$.  This is a 1.2\% correction to the \pp and \raa results.

Collision events are required to have at least one primary vertex~\cite{CMS:2014pgm} formed from two tracks within 15\unit{cm} of the detector center in the $z$ direction.  Additionally, \OO events are required to produce energy deposits above 13\GeV in one or more detector elements on each side of the HF.  These requirements result in a selection inefficiency below 1\% for events producing a charged particle having $\pt>3\GeV$ and pseudorapidity $\abs{\eta}<1$, and a correction for the remaining inefficiency is applied.  Similarly, residual contamination from ultraperipheral events~\cite{Baltz:2007kq} is found to be negligible for this measurement.  Ion beam contamination resulting from electromagnetic dissociation processes~\cite{Pshenichnov:1997un} of $^{16}$O was found to be negligible for this analysis by studying the time-dependence of event activity distributions~\cite{Das:2025rqb}.

The measurements reported here are for charged primary particles, which are defined as all charged particles having an average proper lifetime $c\tau>1\unit{cm}$.  Particles that result from secondary decays are not considered primary if the parent particle is classified as primary.  Charged particles are examined using the hits they leave in the silicon tracker.  Tracks and the vertices from which they originate are reconstructed using algorithms discussed in Ref.~\cite{CMS:2014pgm}.  Tracks must satisfy the ``high purity'' selection described therein to reduce contamination resulting from spurious tracks not associated with a charged particle, also known as misreconstruction.  Only tracks in the ranges $\abs{\eta}<1$ and $3.0<\pt<103.6\GeV$ are used. If a track has $\pt>10\GeV$, the relative uncertainty in its \pt is required to be less than 10\%, as determined by the track fit.  The misreconstruction rate of tracks is found to be below 1\% in simulation. To enhance the fraction of tracks that correspond to primary particles, the analysis only considers tracks having a distance of closest approach (DCA) to any primary vertex less than three times the associated DCA uncertainty in both the $x$-$y$ plane and $z$ direction. In simulation, contamination effects from secondary particles are around 0.5\% after these selections.  To facilitate cancellation of reconstruction effects and uncertainties when formulating the \raa observable, identical track reconstruction procedures and selections are applied to both \pp and \OO datasets.

The resulting tracking efficiency is evaluated using the \PYTHIA~\cite{Bierlich:2022pfr} generator for \pp collisions and using \PYTHIA overlaid with \HIJING~\cite{Wang:1991hta} for \OO collisions.  For a majority of particles considered ($\PGppm$, $\PKpm$, $\Pp$, and $\PAp$), the efficiency is between 89\% and 91\% in both collision systems. Charged strange baryons ($\PGSpm$, $\PGXm$, $\PGOm$, and their antiparticles) have lifetimes long enough to be considered primary particles, but mostly decay before traversing enough of the tracker to be reconstructed.  \PYTHIA underestimates the production rates of strange particles in \pp~\cite{ALICE:2018vuu} and \pPb~\cite{ALICE:2016fzo} collisions, resulting in a slight overestimation of the tracking efficiency.  To account for this effect, the relative production rates of different particle types in simulation are weighted to match those in 7\TeV \pp data~\cite{ALICE:2020jsh,ALICE:2016dei} using a procedure similar to the one outlined in Ref.~\cite{ALICE:2018vuu}.  In \pp collisions, this reduces the overall charged-particle reconstruction efficiency calculated using \PYTHIA by around 6\% at $\pt=3\GeV$ but results in a less than 1\% difference for $\pt>8\GeV$. For \OO collisions, a similar procedure is used with an additional extrapolation factor, based on control samples in data, that accounts for expected differences between the \OO and \pp data.  Motivated by measurements~\cite{ALICE:2019hno,ALICE:2016fzo} indicating that the hadron composition of an event scales in a universal way with charged-particle density, \dNdEta, this factor estimates the change of light-hadron production rates with event activity. Specifically, the particle abundances used to weight the simulation are scaled in a \pt-independent manner from a \pp-like working point of $\dNdEta=8$ to the measured value for inclusive \OO collisions of $\dNdEta\approx40$~\cite{CMS-PAS-HIN-25-010}.  This amounts to an additional 3\% correction at $\pt=3\GeV$ which decreases to below 1\% for $\pt>10\GeV$.  To compensate for track reconstruction inefficiencies, residual track misreconstruction effects, and secondary-track contamination, correction factors are applied on a track-by-track basis as functions of \pt and $\eta$.  

The uncertainty in the tracking efficiency due to the particle composition weighting procedure has two components.  The first is calculated by propagating the experimental uncertainties in the input \pp data through the entire procedure.  This uncertainty is less than 2.0\% for $\pt>10\GeV$ in the \pp and \OO cross sections but fully cancels in \raa.  The second component results from the extrapolation procedure used to account for differences between \OO and \pp data.  This component is calculated by examining two variations: one with no extrapolation and one where the extrapolation goes to $\dNdEta=100$, corresponding to the density expected in the most head-on \OO collisions. This uncertainty, which is only applicable to the \OO spectra and \raa measurements, is 3\% at $\pt=3\GeV$ and less than 1\% at $\pt>10\GeV$.
The uncertainty caused by potential mismodeling of the tracking efficiency in simulation (other than the previously discussed differences in the types of particles produced) is evaluated by comparing the relative reconstruction efficiencies of pions coming from different $\PDz$ meson decays~\cite{CMS-PAS-TRK-10-002} in 13\TeV \pp data.  This uncertainty is 2.1\% for both \pp and \OO collisions and cancels fully when calculating \raa.   To study the sensitivity of the measurement to secondary-track backgrounds, the analysis is repeated after changing the DCA selection to require tracks to have less than two or five times their DCA uncertainty.  The maximum of the two variations is taken as a systematic uncertainty and is up to 3.5\% in \pp and 5.6\% in \OO collisions.  This uncertainty is under 3.5\% for \raa because of partial cancellations.  The \pp pileup correction factor of 1.2\% is taken as an uncertainty for the \pp spectrum and \raa measurements.  The uncertainty attributed to bin migration effects caused by finite track \pt resolution is less than 1\% for the spectra measurements and fully cancels in the \raa measurement.  Uncertainties resulting from the luminosity calibrations are not included in the total systematic uncertainty but are reported as a separate normalization uncertainty that is fully correlated across all \pt values probed.  When compared to the \pp spectrum measured at 5.02\TeV by CMS in Ref.~\cite{CMS:2016xef}, the precision achieved for the 5.36\TeV \pp measurement improved from around 7\% to about 4\% because of a better understanding of the tracking efficiency, a more sophisticated treatment of pileup effects, and new constraints on the particle composition weighting procedure.  Summaries of the systematic uncertainties are provided in \addmatter{endmatter:suppMat2}.

\begin{figure}[ht]
    \centering
    \includegraphics[width=\cmsFigWidth]{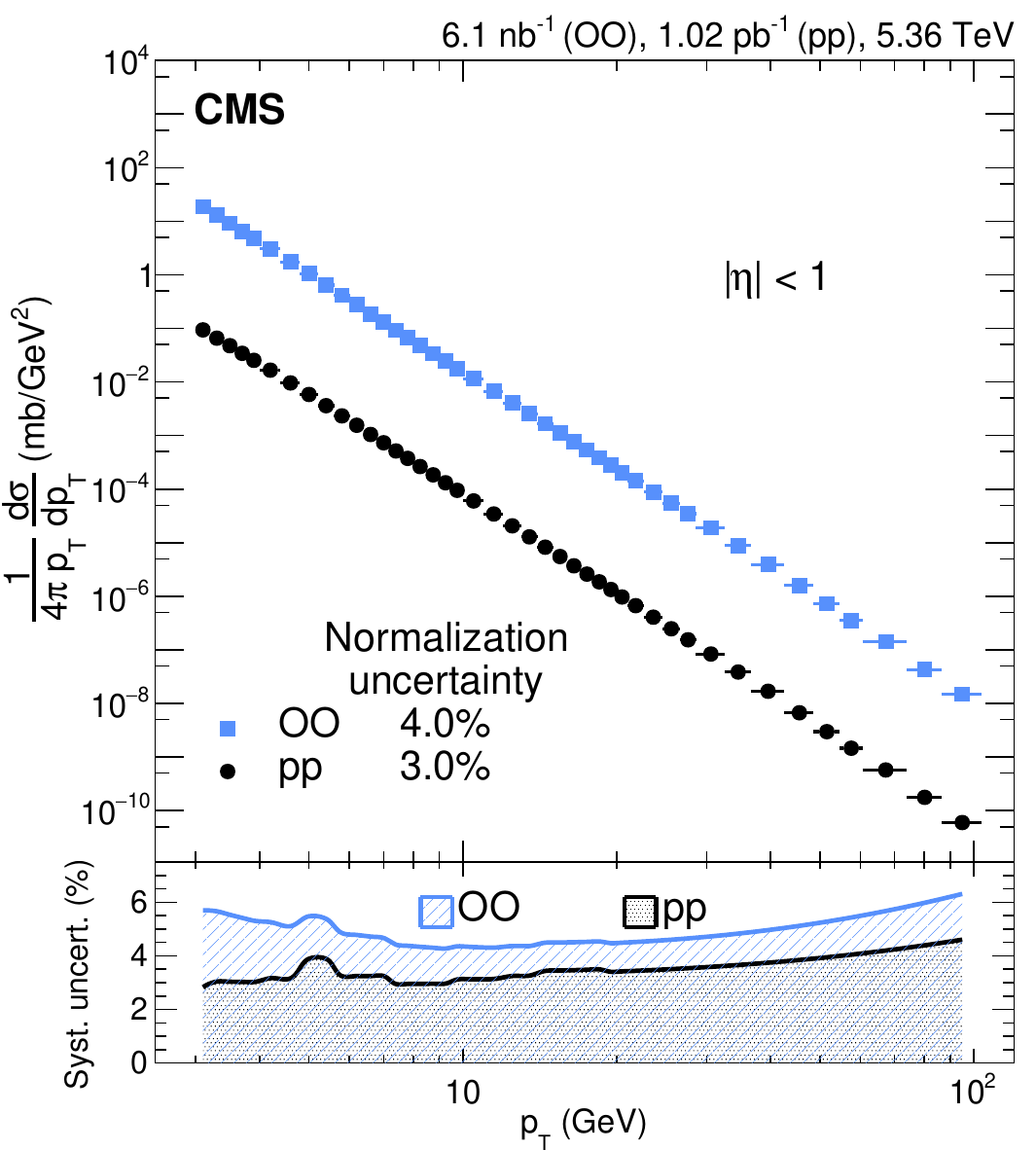}
    \caption{The charged-particle \pt-differential invariant cross sections for \pp (black circles) and \OO  (blue squares) collisions.  The markers show the average cross section across the entire bin width, not the cross section value at the center of each bin.  Statistical uncertainties are smaller than the markers.}
    \label{fig:Spectra}
\end{figure}

The \pt-differential invariant cross sections of charged hadrons are reported in Fig.~\ref{fig:Spectra}.  Both spectra exhibit roughly power law behavior, which is consistent with previous measurements of charged-particle spectra at these energies~\cite{CMS:2016xef, CMS:2015ved, CMS:2011mry, CMS:2018nhd, CMS:2018yyx}.
\begin{figure}[ht]
    \centering
    \includegraphics[width=\cmsFigWidth]{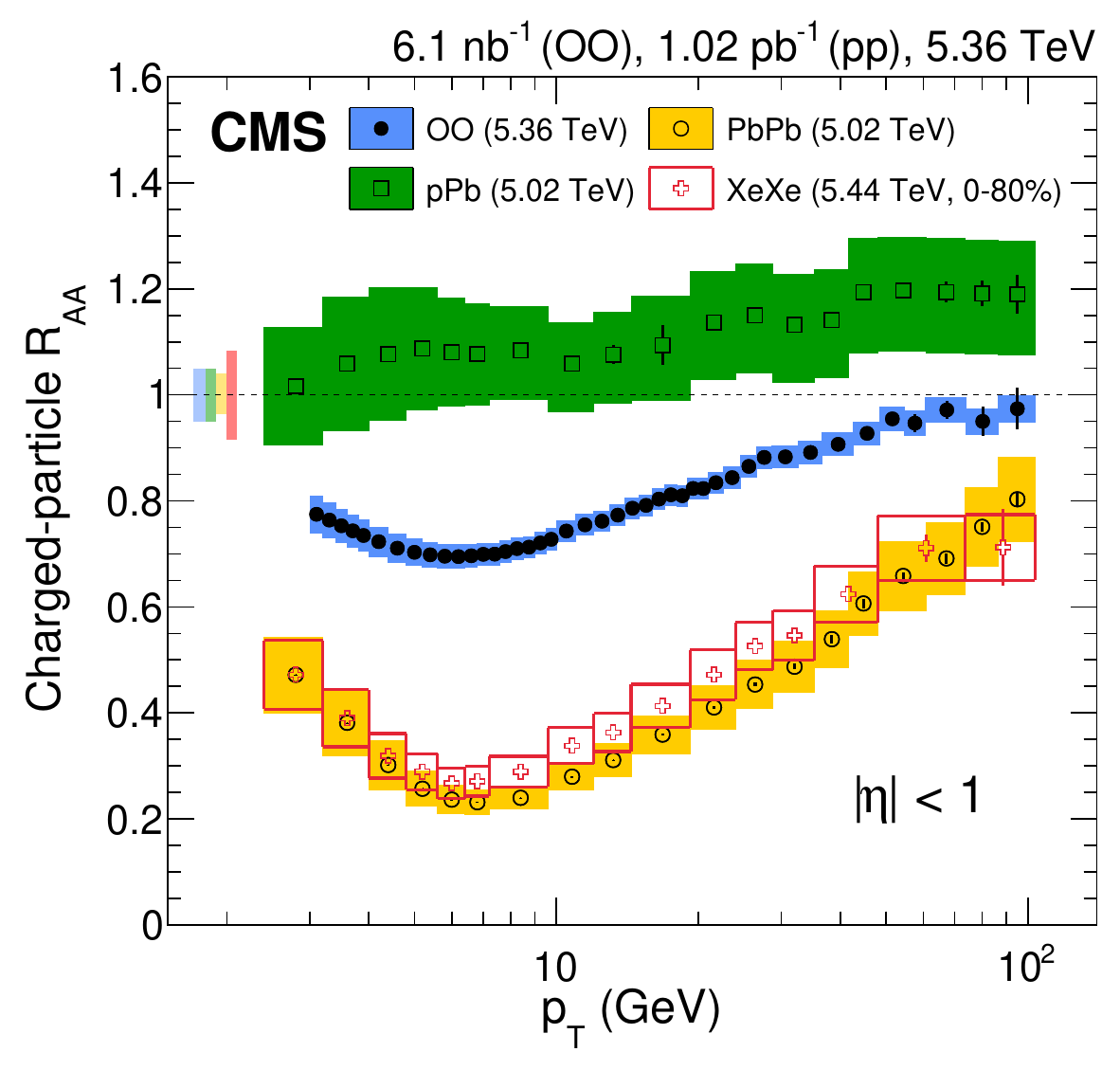}
    \caption{The charged-particle \raa for inclusive 5.36 \TeV \OO collisions (solid markers). Error bars represent statistical uncertainties and boxes represent systematic uncertainties. The \raa for \pPb~\cite{CMS:2016xef,CMS:2015ved}, \XeXe~\cite{CMS:2018yyx}, and \PbPb~\cite{CMS:2016xef} collisions (open markers) are also shown.  Normalization uncertainties are shown by the bands around unity. 
 }
    \label{fig:RAA}
\end{figure}

The measured \raa, shown in Fig.~\ref{fig:RAA}, exhibits significant \pt dependence with the lowest \pt data having \raa around 0.77.  As the probed \pt increases, the \raa  exhibits a local minimum at around $\pt=6\GeV$ before increasing to a value consistent with unity at \pt values approaching 100\GeV.  The \raa value of $0.69\pm0.04$ (systematic + normalization, with negligible statistical uncertainty) at the local minimum is more than five standard deviations from unity, making this the first definitive observation of particle suppression in \OO collisions.  For comparison, previous CMS measurements of the inclusive charged-particle \raa are shown for \pPb and PbPb collisions at $\sqrtsNN=5.02\TeV$~\cite{CMS:2016xef,CMS:2015ved}, as well as the \raa for XeXe data at $\sqrtsNN=5.44\TeV$~\cite{CMS:2018yyx}.  The \XeXe sample corresponds to the top 80\% most head-on collisions, which is expected to be close to the \raa values for an inclusive event selection.  The behavior of the \OO \raa is qualitatively similar to that observed in larger collision systems such as \XeXe and \PbPb.  In those systems, the downward slope at lower \pt is typically attributed to the \pt-dependence of effects such as Cronin enhancement~\cite{Antreasyan:1978cw} and radial flow~\cite{ATLAS:2025ztg, ALICE:2025iud}, whereas the magnitude and trend of suppression at higher \pt is associated with the strength of parton energy loss.  An ordering with collision system size is apparent, with the \raa for \pPb collisions being above, and for \XeXe and \PbPb collisions being below the \OO data.

\begin{figure*}[ht]
    \centering
    \includegraphics[width=\textwidth]{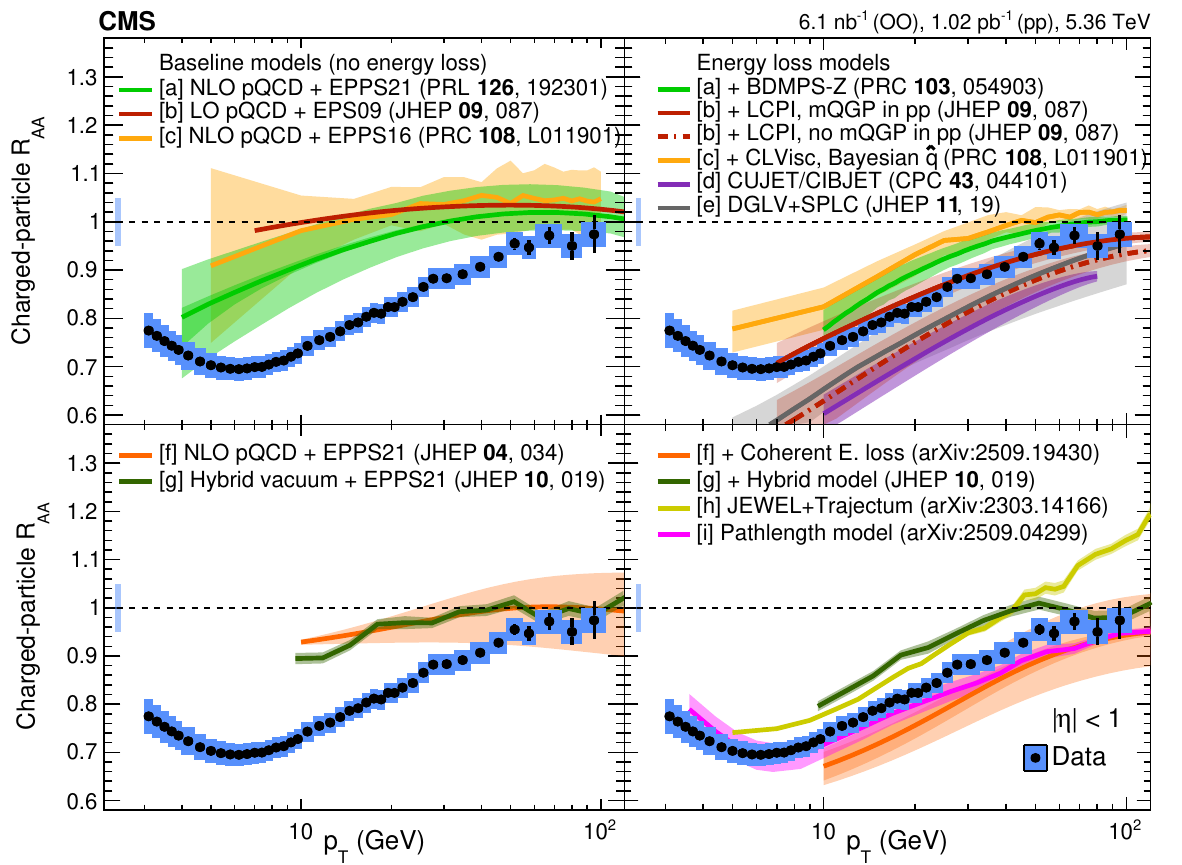}
    \caption{Comparison of the OO \raa measurement to various theoretical predictions~\cite{Huss:2020dwe,Huss:2020whe,Zakharov:2021uza,Xie:2022ght,Shi:2018izg,Shi:2018lsf,Casalderrey-Solana:2014bpa, Gebhard:2024flv, Pablos:2025cli, vanderSchee:2025hoe,Faraday:2025pto,Mazeliauskas:2025clt}.  The normalization uncertainty of the measurement is shown by the band around unity on the left side of each panel.  The uncertainty bands around the predictions indicate 68\% confidence limit intervals, except for model [c], where they indicate a 90\% interval.}
    \label{fig:theory}
\end{figure*}

Figure~\ref{fig:theory} compares the \OO \raa to theoretical predictions.  Predictions in the left panels do not include parton energy loss, while those in the right panels include energy loss.   Model [a] is a next-to-leading order (NLO) perturbative QCD (pQCD) calculation~\cite{Mazeliauskas:2025clt,Huss:2020dwe} interfaced with the EPPS21 nuclear parton distribution functions (nPDFs)~\cite{Eskola:2021nhw}.  The outer band represents nPDF uncertainties, while the inner band shows scale uncertainties of the calculation.   Comparisons to this model with other nPDF choices are shown in \suppmat{app:suppMat3}\ifthenelse{\boolean{cms@external}}{, which contains Refs.~\cite{AbdulKhalek:2022fyi,Helenius:2021tof,Klasen:2023uqj,Paakkinen:2021jjp}}{}.  The model in the left panel does not include parton energy loss effects, but it is interfaced with a parton energy loss model using the BDMPS-Z framework~\cite{Huss:2020whe} in the right panel.  Some suppression of the \raa, particularly in the range of $\pt<20\GeV$, can clearly be attributed to the baseline model, but the inclusion of energy loss effects brings the prediction closer to the data.  Model [b] is a leading order calculation~\cite{Zakharov:2021uza} interfaced with the EPS09~\cite{Eskola:2009uj} nPDF set.  In the right panel, it is combined with the LCPI energy loss framework~\cite{Zakharov:2021uza} in two scenarios where a QGP droplet, denoted as ``mQGP", is or is not created in \pp collisions.  The inclusion of energy loss effects clearly improves the agreement with the data, with the measurement favoring the scenario where a QGP droplet is formed in \pp collisions.  Model [c] represents a NLO pQCD model using the EPPS16~\cite{Eskola:2016oht} nPDF set.  In the right panel, it is combined with the CLVisc (3+1)D hydrodynamics model~\cite{Pang:2012he}, with energy loss included by incorporating a jet transport coefficient $\hat{q}$ determined using Bayesian inference techniques~\cite{Xie:2022ght}.  The uncertainty band results from variations in $\hat{q}$.  This model seems to qualitatively capture the shape of the data but predicts a smaller suppression magnitude. Models [d] and [e] do not include nPDF effects, \ie, the baseline equals unity.   Model [d] illustrates the CUJET/CIBJET model~\cite{Shi:2018izg,Shi:2018lsf}, which combines viscous hydrodynamics with a DGLV theory of energy loss~\cite{Djordjevic:2003zk,Gyulassy:1999zd}.  The uncertainty band reflects variations in the strong coupling constant used.  Model [e] is similar and utilizes DGLV radiative energy loss combined with collisional energy loss and an additional short path length correction (SPLC) for small systems~\cite{Faraday:2025pto}.  The uncertainty reflects variations induced by the choice of the cutoff momentum scales used.  Both [d] and [e] predict smaller \raa compared to data, and [e] is consistent within data within uncertainties for $\pt > 20\GeV$.  Model [f] is a NLO pQCD baseline model~\cite{Gebhard:2024flv} interfaced with EPPS21, and the width of the corresponding band shows nPDF and scale uncertainties.  Energy loss effects are added in the right panel by coupling the model to event-by-event hydrodynamics simulations and including coherent energy loss~\cite{Pablos:2025cli,Mehtar-Tani:2021fud,Takacs:2021bpv,Mehtar-Tani:2024jtd}.  The darker band represents an additional uncertainty obtained by varying the model parameters after they were constrained using a Bayesian fit to previously-measured hadron spectra. The prediction is consistent with the data within the uncertainties, although its central values illustrate a slightly more pronounced \pt-dependence.  Model [g] represents a baseline calculation that uses EPPS21.  On the right, the Hybrid model~\cite{Casalderrey-Solana:2014bpa} has been used to add energy loss effects using event-by-event hydrodynamics profiles~\cite{Mantysaari:2025tcg}.  Although the presence of energy loss effects causes a suppression of the \raa, the predicted values are still above the data for most of the \pt range examined.  Model [h] is the \textsc{JEWEL}~\cite{Zapp:2013vla} MC generator which has been interfaced with the \textsc{Trajectum} code~\cite{Kolbe:2023rsq,Nijs:2020ors, Nijs:2020roc, vanderSchee:2025hoe}.  It predicts a larger \raa value as compared to the data, with the values exceeding unity at high \pt.  The \textsc{JEWEL} generator is expected to perform better when examining jets rather than individual hadrons, which could explain this behavior.  Note that only statistical uncertainties are shown for the Hybrid model, its baseline, and \textsc{JEWEL}+\textsc{Trajectum}, while other uncertainties were neglected.  Model [i] extrapolates from \PbPb \raa values using a path length integral and the temperature profile experienced by partons as they travel through the medium, as calculated using \textsc{Trajectum}~\cite{Nijs:2020ors, Nijs:2020roc, vanderSchee:2023uii,vanderSchee:2025hoe}.  It describes the data well.  

The comparisons with models reveal conclusions that appear to be independent of the exact details of the model examined.  Firstly, the role of nPDFs is clearly important---the oxygen nPDF alone can potentially cause an \raa suppression up to half of that seen in data.  Further nPDF constraints coming from the recently-collected LHC \OO and $\Pp$O datasets may improve this in the future.  Secondly, models including energy loss effects describe \raa better than baseline models.  Most models capture the \pt-dependent shape of \raa for $\pt>10\GeV$, but a sizable spread exists in the predicted scale of \raa.  Overall, the data are consistent with the presence of parton energy loss effects in \OO collisions, and this measurement places quantitative constraints on models incorporating energy loss in small collision systems.

In summary, we report the first measurement of the charged-particle nuclear modification factor (\raa) as a function of transverse momentum (\pt) in oxygen-oxygen collisions at a center of mass energy per nucleon pair of 5.36\TeV.  The \raa is significantly suppressed and reaches a minimum of $0.69\pm0.04$ around $\pt=6\GeV$.  However, \raa rises to become consistent with unity for \pt values approaching 100 \GeV.  Comparisons with previous measurements of other collision systems reveal a system-size dependence of \raa suppression in nuclear collisions.  The data are more consistent with theoretical models that incorporate parton energy loss effects as compared to baseline models without energy loss, supporting the hypothesis that a medium of sufficient size to give rise to observable parton energy loss is created.  This measurement constrains models attempting to predict jet quenching in small systems and represents an important first step towards the realization of a physics program examining collisions of light nuclei at \TeV-scale energies.

\begin{acknowledgments}
We congratulate our colleagues in the CERN accelerator departments for the excellent performance of the LHC and thank the technical and administrative staffs at CERN and at other CMS institutes for their contributions to the success of the CMS effort. In addition, we gratefully acknowledge the computing centers and personnel of the Worldwide LHC Computing Grid and other centers for delivering so effectively the computing infrastructure essential to our analyses. Finally, we acknowledge the enduring support for the construction and operation of the LHC, the CMS detector, and the supporting computing infrastructure provided by the following funding agencies: SC (Armenia), BMBWF and FWF (Austria); FNRS and FWO (Belgium); CNPq, CAPES, FAPERJ, FAPERGS, and FAPESP (Brazil); MES and BNSF (Bulgaria); CERN; CAS, MoST, and NSFC (China); MINCIENCIAS (Colombia); MSES and CSF (Croatia); RIF (Cyprus); SENESCYT (Ecuador); ERC PRG, TARISTU24-TK10 and MoER TK202 (Estonia); Academy of Finland, MEC, and HIP (Finland); CEA and CNRS/IN2P3 (France); SRNSF (Georgia); BMFTR, DFG, and HGF (Germany); GSRI (Greece); NKFIH (Hungary); DAE and DST (India); IPM (Iran); SFI (Ireland); INFN (Italy); MSIT and NRF (Republic of Korea); MES (Latvia); LMTLT (Lithuania); MOE and UM (Malaysia); BUAP, CINVESTAV, CONACYT, LNS, SEP, and UASLP-FAI (Mexico); MOS (Montenegro); MBIE (New Zealand); PAEC (Pakistan); MES, NSC, and NAWA (Poland); FCT (Portugal); MESTD (Serbia); MICIU/AEI and PCTI (Spain); MOSTR (Sri Lanka); Swiss Funding Agencies (Switzerland); MST (Taipei); MHESI (Thailand); TUBITAK and TENMAK (T\"{u}rkiye); NASU (Ukraine); STFC (United Kingdom); DOE and NSF (USA).
\end{acknowledgments}

\bibliography{auto_generated} 

\appendix
\numberwithin{figure}{section}
\ifthenelse{\boolean{cms@external}}{\section{}~\nocite{AbdulKhalek:2022fyi,Helenius:2021tof,Klasen:2023uqj,Paakkinen:2021jjp}}{\section{Additional material}\label{endmatter}}
\subsection{Description of event simulations}
\label{endmatter:suppMat1}
Simulated MC samples are used to evaluate detector performance and calculate correction factors.  For \pp collisions, the \PYTHIA 8.311 event generator~\cite{Bierlich:2022pfr} with the CP5 tune~\cite{CMS:2019csb} is employed.  To study \OO collisions, a \PYTHIA event is overlaid with an \OO~\HIJING v1.383~\cite{Wang:1991hta} event to form a composite event containing both a hard scattering process as well as particle production from the underlying event.  Potential contamination from ultraperipheral collisions~\cite{Baltz:2007kq} is studied using \textsc{STARlight} v3.3.0~\cite{Klein:2016yzr}.  The \GEANTfour~\cite{Agostinelli:2002hh} program is used to simulate the CMS detector response.  After event generation, an event-by-event weight is used to ensure that the distribution of the $z$ position of reconstructed primary interaction points in simulation matches that in data.

\subsection{Summaries of the uncertainties}
\label{endmatter:suppMat2}

\begin{figure}[th!]
    \centering
      \includegraphics[width=.49\textwidth]{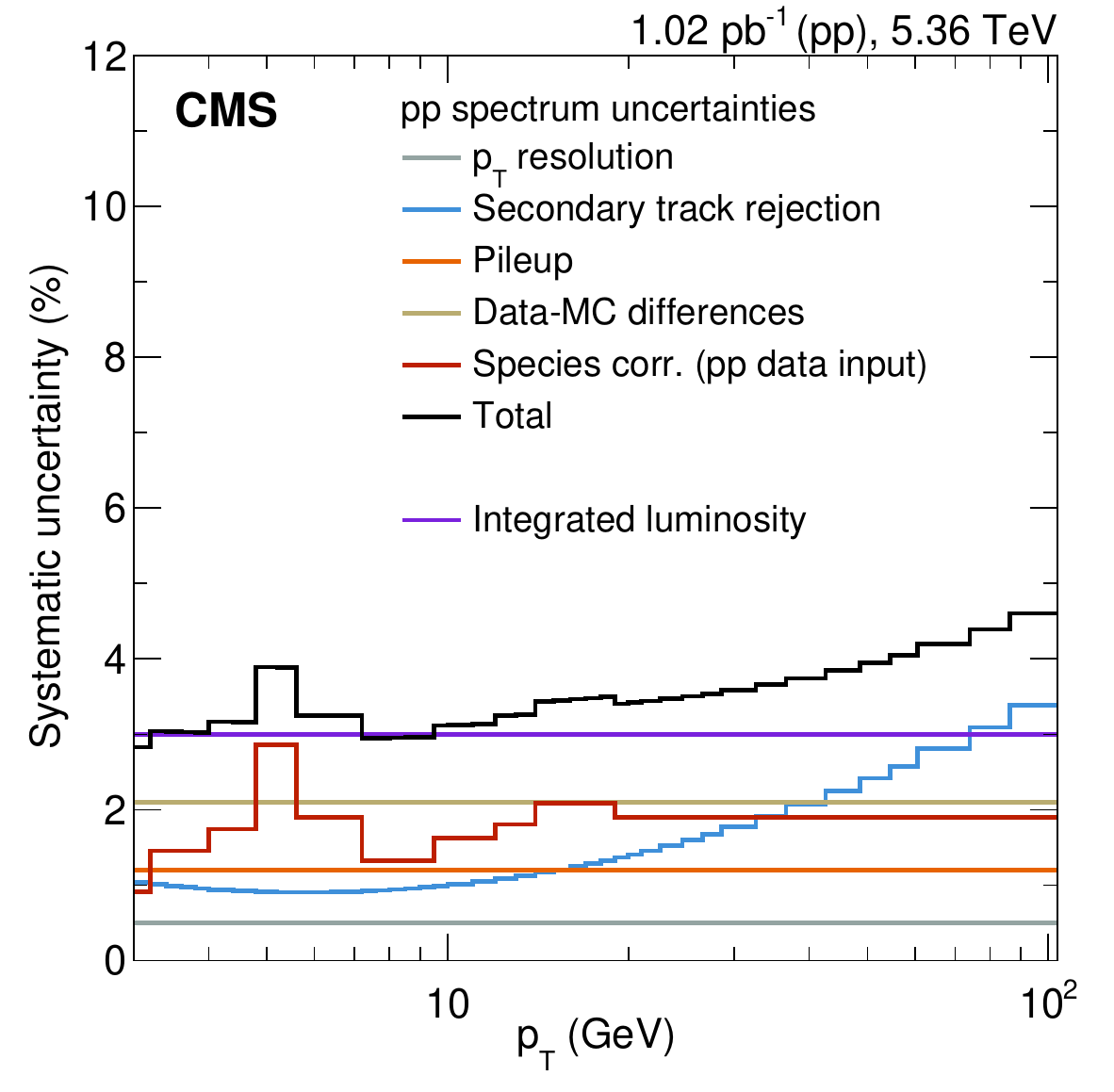}
      \includegraphics[width=.49\textwidth]{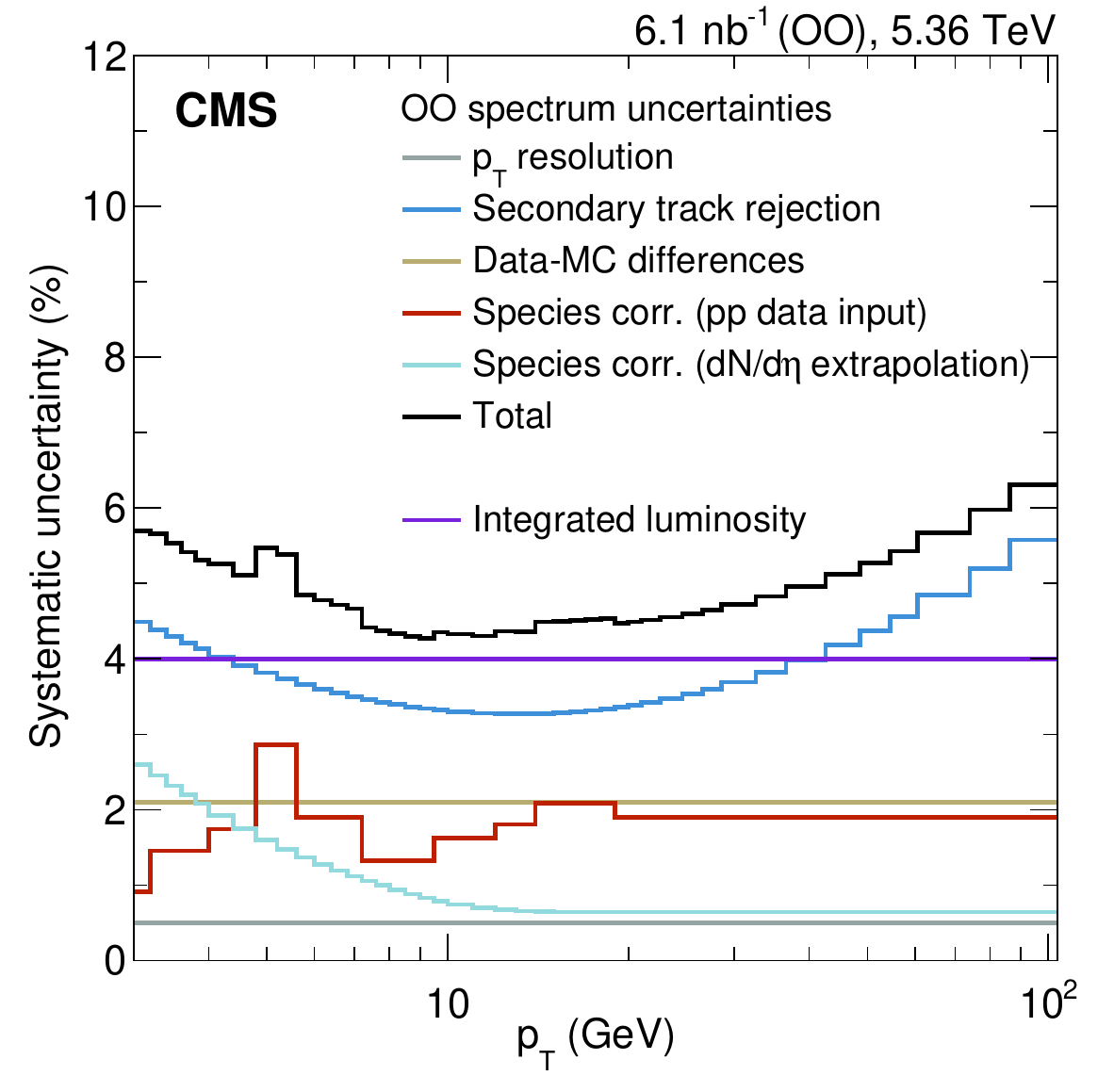}
    \caption{Systematic uncertainties related to the 5.36\TeV \pp (\cmsLeft) and \OO (\cmsRight) cross section measurements, presented as a function of \pt. }
    \label{endmatter:SpecUncert}
\end{figure}

\begin{figure}[th!]
    \centering
    \includegraphics[width=\cmsFigWidth]{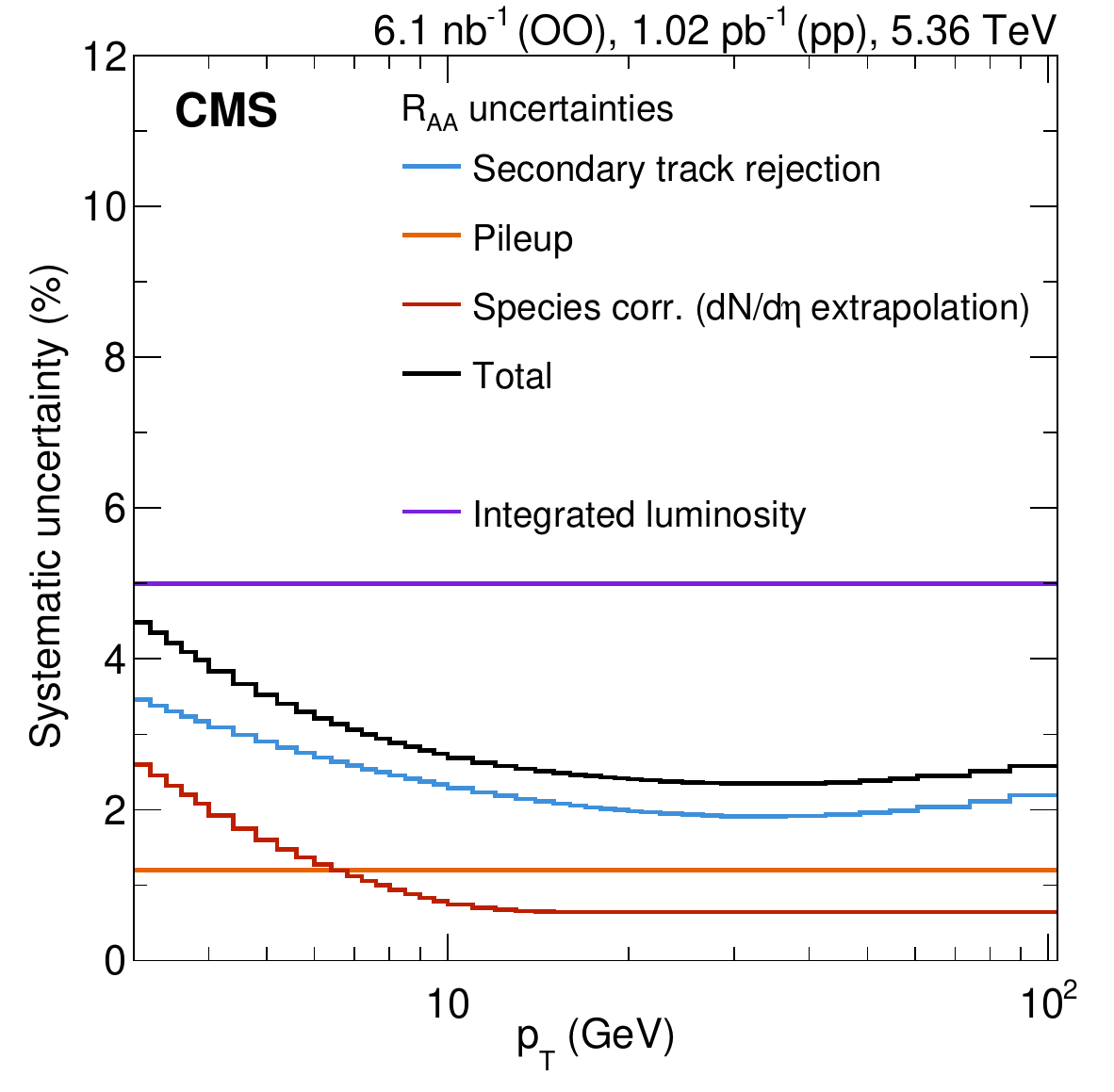}
    \caption{Systematic uncertainties related to the measurement of \raa for 5.36\TeV \OO collisions, presented as a function of \pt. }
    \label{endmatter:RAAUncert}
\end{figure}

A visual summary of the systematic uncertainties relevant for the spectra measured in this analysis is shown in Fig.~\ref{endmatter:SpecUncert}.  The integrated luminosity uncertainty is reported as a separate uncertainty and is not included in the total coming from other sources.  The non-monotonic behavior at lower \pt values results from propagating the experimental uncertainties of the input ALICE data used for the particle species correction. Figure~\ref{endmatter:RAAUncert} shows a similar summary for the \raa measurement after correlations between the \OO and \pp measurements are taken into account.

\ifthenelse{\boolean{cms@external}}{}{
\input{supplemental_material}

}

\cleardoublepage \section{The CMS Collaboration \label{app:collab}}\begin{sloppypar}\hyphenpenalty=5000\widowpenalty=500\clubpenalty=5000\input{HIN-25-008-public-authorlist.tex}\end{sloppypar}
\end{document}

%% file: supplemental_material.tex
\subsection{Comparison to different nPDF models}
\label{app:suppMat3}

\begin{figure}[th]
    \centering
    \ifthenelse{\boolean{cms@external}}{\includegraphics[width=.95\columnwidth]{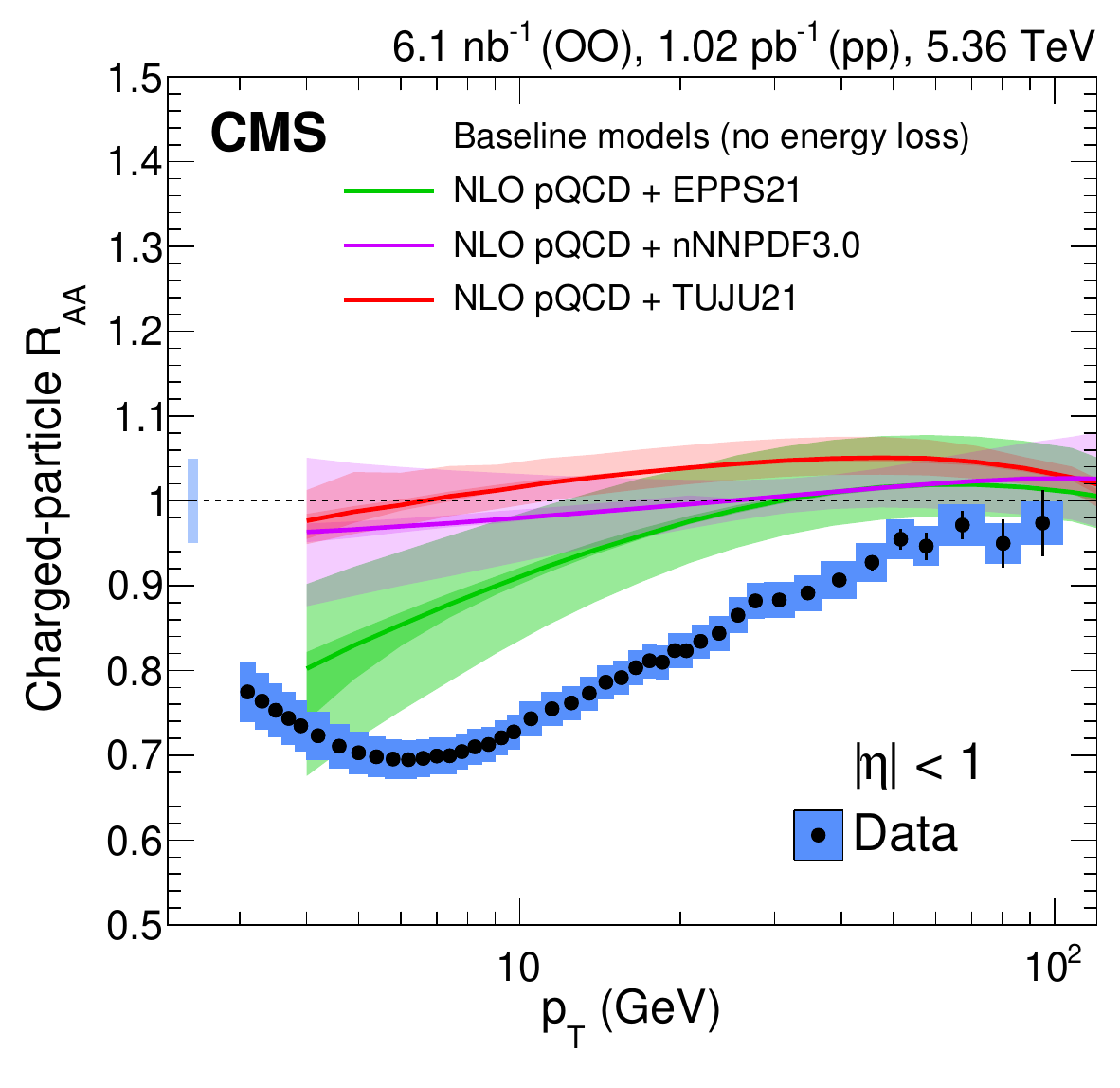}}{\includegraphics[width=.47\linewidth]{Figure_006.pdf}}
    \caption{Comparison of the \OO \raa to a theoretical model~\cite{Mazeliauskas:2025clt,Huss:2020dwe} without energy loss effects that uses three different nPDF choices~\cite{Eskola:2021nhw,AbdulKhalek:2022fyi,Helenius:2021tof}.}
    \label{fig:nPDFs}
\end{figure}

Figure~\ref{fig:nPDFs} shows the baseline model described in Refs.~\cite{Mazeliauskas:2025clt,Huss:2020dwe} with three different nPDF hypotheses.  The data lie under all three nPDF choices.  The EPPS21 model predicts the strongest nPDF modification of \raa, whereas the nNNPDF3.0~\cite{AbdulKhalek:2022fyi} and TUJU21~\cite{Helenius:2021tof} nPDF sets predict a rather modest suppression effect.  TUJU21 currently does not include LHC data having hadronic final states.  These data tend to favor a stronger shadowing of the gluon nPDF, so this difference could potentially explain why TUJU21 and EPPS21 predict different behavior~\cite{Klasen:2023uqj}.  EPPS21 and nNNPDF3.0 include similar data in their global fits, but the models predict different A-dependence of nPDF modifications.  In particular, nuclear effects tend to be smaller for low-A nuclei in nNNPDF3.0 as compared to EPPS21~\cite{Paakkinen:2021jjp}.

%% file: HIN-25-008-public-authorlist.tex
\cmsinstitute{Yerevan Physics Institute, Yerevan, Armenia}
{\tolerance=6000
A.~Hayrapetyan, V.~Makarenko\cmsorcid{0000-0002-8406-8605}, A.~Tumasyan\cmsAuthorMark{1}\cmsorcid{0009-0000-0684-6742}
\par}
\cmsinstitute{Institut f\"{u}r Hochenergiephysik, Vienna, Austria}
{\tolerance=6000
W.~Adam\cmsorcid{0000-0001-9099-4341}, L.~Benato\cmsorcid{0000-0001-5135-7489}, T.~Bergauer\cmsorcid{0000-0002-5786-0293}, M.~Dragicevic\cmsorcid{0000-0003-1967-6783}, C.~Giordano\cmsorcid{0000-0001-6317-2481}, P.S.~Hussain\cmsorcid{0000-0002-4825-5278}, M.~Jeitler\cmsAuthorMark{2}\cmsorcid{0000-0002-5141-9560}, N.~Krammer\cmsorcid{0000-0002-0548-0985}, A.~Li\cmsorcid{0000-0002-4547-116X}, D.~Liko\cmsorcid{0000-0002-3380-473X}, M.~Matthewman, J.~Schieck\cmsAuthorMark{2}\cmsorcid{0000-0002-1058-8093}, R.~Sch\"{o}fbeck\cmsAuthorMark{2}\cmsorcid{0000-0002-2332-8784}, M.~Shooshtari\cmsorcid{0009-0004-8882-4887}, M.~Sonawane\cmsorcid{0000-0003-0510-7010}, W.~Waltenberger\cmsorcid{0000-0002-6215-7228}, C.-E.~Wulz\cmsAuthorMark{2}\cmsorcid{0000-0001-9226-5812}
% ...
\par}
\vskip\cmsinstskip
\dag:~Deceased\\
$^{1}$Also at Yerevan State University, Yerevan, Armenia\\
$^{2}$Also at TU Wien, Vienna, Austria\\
% ...


%% file: HIN-25-008_temp.bbl
\providecommand{\href}[2]{#2}\begingroup\raggedright\begin{thebibliography}{10}%
\makeatletter
\providecommand{\hrefCMSnoop }[0]{\@secondoftwo}%
\makeatother
\providecommand{\doi}{\texttt{doi:}\begingroup \urlstyle{tt}\Url}

\bibitem{Busza:2018rrf}
\hrefCMSnoop {}{W.~Busza, K.~Rajagopal, and W.~van~der Schee, ``{Heavy ion
  collisions: The big picture, and the big questions}'',} \textit{ Ann. Rev.
  Nucl. Part. Sci.} \textbf{ 68} (2018) 339,
  \href{http://dx.doi.org/10.1146/annurev-nucl-101917-020852}{\doi{10.1146/annurev-nucl-101917-020852}},
  \href{http://www.arXiv.org/abs/1802.04801}{\texttt{arXiv:1802.04801}}.

\bibitem{Shuryak:1980tp}
\hrefCMSnoop {}{E.~V. Shuryak, ``{Quantum chromodynamics and the theory of
  superdense matter}'',} \textit{ Phys. Rept.} \textbf{ 61} (1980) 71,
  \href{http://dx.doi.org/10.1016/0370-1573(80)90105-2}{\doi{10.1016/0370-1573(80)90105-2}}.

\bibitem{CMS:2024krd}
\hrefCMSnoop {}{{CMS Collaboration}, ``{Overview of high-density QCD studies
  with the CMS experiment at the LHC}'',} \textit{ Phys. Rept.} \textbf{ 1115}
  (2025) 219,
  \href{http://dx.doi.org/10.1016/j.physrep.2024.11.007}{\doi{10.1016/j.physrep.2024.11.007}},
  \href{http://www.arXiv.org/abs/2405.10785}{\texttt{arXiv:2405.10785}}.

\bibitem{ALICE:2022wpn}
\hrefCMSnoop {}{{ALICE Collaboration}, ``{The ALICE experiment: a journey
  through QCD}'',} \textit{ Eur. Phys. J. C} \textbf{ 83} (2024) 813,
  \href{http://dx.doi.org/10.1140/epjc/s10052-024-12935-y}{\doi{10.1140/epjc/s10052-024-12935-y}},
  \href{http://www.arXiv.org/abs/2211.04384}{\texttt{arXiv:2211.04384}}.

\bibitem{BRAHMS:2004adc}
\hrefCMSnoop {}{{BRAHMS} Collaboration, ``{Quark gluon plasma and color glass
  condensate at RHIC? The perspective from the BRAHMS experiment}'',} \textit{
  Nucl. Phys. A} \textbf{ 757} (2005) 1,
  \href{http://dx.doi.org/10.1016/j.nuclphysa.2005.02.130}{\doi{10.1016/j.nuclphysa.2005.02.130}},
  \href{http://www.arXiv.org/abs/nucl-ex/0410020}{\texttt{arXiv:nucl-ex/0410020}}.

\bibitem{PHOBOS:2004zne}
\hrefCMSnoop {}{{PHOBOS} Collaboration, ``{The PHOBOS perspective on
  discoveries at RHIC}'',} \textit{ Nucl. Phys. A} \textbf{ 757} (2005) 28,
  \href{http://dx.doi.org/10.1016/j.nuclphysa.2005.03.084}{\doi{10.1016/j.nuclphysa.2005.03.084}},
  \href{http://www.arXiv.org/abs/nucl-ex/0410022}{\texttt{arXiv:nucl-ex/0410022}}.

\bibitem{STAR:2005gfr}
\hrefCMSnoop {}{{STAR} Collaboration, ``{Experimental and theoretical
  challenges in the search for the quark gluon plasma: The STAR Collaboration's
  critical assessment of the evidence from RHIC collisions}'',} \textit{ Nucl.
  Phys. A} \textbf{ 757} (2005) 102,
  \href{http://dx.doi.org/10.1016/j.nuclphysa.2005.03.085}{\doi{10.1016/j.nuclphysa.2005.03.085}},
  \href{http://www.arXiv.org/abs/nucl-ex/0501009}{\texttt{arXiv:nucl-ex/0501009}}.

\bibitem{PHENIX:2004vcz}
\hrefCMSnoop {}{{PHENIX} Collaboration, ``{Formation of dense partonic matter
  in relativistic nucleus-nucleus collisions at RHIC: Experimental evaluation
  by the PHENIX collaboration}'',} \textit{ Nucl. Phys. A} \textbf{ 757} (2005)
  184,
  \href{http://dx.doi.org/10.1016/j.nuclphysa.2005.03.086}{\doi{10.1016/j.nuclphysa.2005.03.086}},
  \href{http://www.arXiv.org/abs/nucl-ex/0410003}{\texttt{arXiv:nucl-ex/0410003}}.

\bibitem{Gross:1973id}
\hrefCMSnoop {}{D.~J. Gross and F.~Wilczek, ``{Ultraviolet behavior of
  nonabelian gauge theories}'',} \textit{ Phys. Rev. Lett.} \textbf{ 30} (1973)
  1343,
  \href{http://dx.doi.org/10.1103/PhysRevLett.30.1343}{\doi{10.1103/PhysRevLett.30.1343}}.

\bibitem{Politzer:1973fx}
\hrefCMSnoop {}{H.~D. Politzer, ``{Reliable perturbative results for strong
  interactions?}'',} \textit{ Phys. Rev. Lett.} \textbf{ 30} (1973) 1346,
  \href{http://dx.doi.org/10.1103/PhysRevLett.30.1346}{\doi{10.1103/PhysRevLett.30.1346}}.

\bibitem{Gyulassy:1990ye}
\hrefCMSnoop {}{M.~Gyulassy and M.~Plumer, ``{Jet quenching in dense
  matter}'',} \textit{ Phys. Lett. B} \textbf{ 243} (1990) 432,
  \href{http://dx.doi.org/10.1016/0370-2693(90)91409-5}{\doi{10.1016/0370-2693(90)91409-5}}.

\bibitem{Cao:2020wlm}
\hrefCMSnoop {}{S.~Cao and X.-N. Wang, ``{Jet quenching and medium response in
  high-energy heavy-ion collisions: a review}'',} \textit{ Rept. Prog. Phys.}
  \textbf{ 84} (2021) 024301,
  \href{http://dx.doi.org/10.1088/1361-6633/abc22b}{\doi{10.1088/1361-6633/abc22b}},
  \href{http://www.arXiv.org/abs/2002.04028}{\texttt{arXiv:2002.04028}}.

\bibitem{Connors:2017ptx}
\hrefCMSnoop {}{M.~Connors, C.~Nattrass, R.~Reed, and S.~Salur, ``{Jet
  measurements in heavy ion physics}'',} \textit{ Rev. Mod. Phys.} \textbf{ 90}
  (2018) 025005,
  \href{http://dx.doi.org/10.1103/RevModPhys.90.025005}{\doi{10.1103/RevModPhys.90.025005}},
  \href{http://www.arXiv.org/abs/1705.01974}{\texttt{arXiv:1705.01974}}.

\bibitem{WA98:2001lnw}
\hrefCMSnoop {}{{WA98} Collaboration, ``{Transverse mass distributions of
  neutral pions from Pb-208 induced reactions at 158-A-GeV}'',} \textit{ Eur.
  Phys. J. C} \textbf{ 23} (2002) 225,
  \href{http://dx.doi.org/10.1007/s100520100886}{\doi{10.1007/s100520100886}},
  \href{http://www.arXiv.org/abs/nucl-ex/0108006}{\texttt{arXiv:nucl-ex/0108006}}.

\bibitem{dEnterria:2004cly}
\hrefCMSnoop {}{D.~G. d'Enterria, ``{Indications of suppressed high p(T) hadron
  production in nucleus - nucleus collisions at CERN-SPS}'',} \textit{ Phys.
  Lett. B} \textbf{ 596} (2004) 32,
  \href{http://dx.doi.org/10.1016/j.physletb.2004.06.071}{\doi{10.1016/j.physletb.2004.06.071}},
  \href{http://www.arXiv.org/abs/nucl-ex/0403055}{\texttt{arXiv:nucl-ex/0403055}}.

\bibitem{CMS:2012aa}
\hrefCMSnoop {}{{CMS Collaboration}, ``{Study of high-\pt charged particle
  suppression in PbPb compared to $\pp$ collisions at $\sqrtsNN=2.76\TeV$}'',}
  \textit{ Eur. Phys. J. C} \textbf{ 72} (2012) 1945,
  \href{http://dx.doi.org/10.1140/epjc/s10052-012-1945-x}{\doi{10.1140/epjc/s10052-012-1945-x}},
  \href{http://www.arXiv.org/abs/1202.2554}{\texttt{arXiv:1202.2554}}.

\bibitem{ALICE:2012aqc}
\hrefCMSnoop {}{{ALICE Collaboration}, ``{Centrality dependence of charged
  particle production at large transverse momentum in Pb--Pb collisions at
  $\sqrtsNN = 2.76\TeV$}'',} \textit{ Phys. Lett. B} \textbf{ 720} (2013) 52,
  \href{http://dx.doi.org/10.1016/j.physletb.2013.01.051}{\doi{10.1016/j.physletb.2013.01.051}},
  \href{http://www.arXiv.org/abs/1208.2711}{\texttt{arXiv:1208.2711}}.

\bibitem{ATLAS:2015qmb}
\hrefCMSnoop {}{{ATLAS Collaboration}, ``{Measurement of charged-particle
  spectra in Pb+Pb collisions at $\sqrtsNN = 2.76\TeV$ with the ATLAS detector
  at the LHC}'',} \textit{ JHEP} \textbf{ 09} (2015) 050,
  \href{http://dx.doi.org/10.1007/JHEP09(2015)050}{\doi{10.1007/JHEP09(2015)050}},
  \href{http://www.arXiv.org/abs/1504.04337}{\texttt{arXiv:1504.04337}}.

\bibitem{CMS:2016xef}
\hrefCMSnoop {}{{CMS Collaboration}, ``{Charged-particle nuclear modification
  factors in PbPb and pPb collisions at $ \sqrtsNN=5.02\TeV$}'',} \textit{
  JHEP} \textbf{ 04} (2017) 039,
  \href{http://dx.doi.org/10.1007/JHEP04(2017)039}{\doi{10.1007/JHEP04(2017)039}},
  \href{http://www.arXiv.org/abs/1611.01664}{\texttt{arXiv:1611.01664}}.

\bibitem{ALICE:2018vuu}
\hrefCMSnoop {}{{ALICE Collaboration}, ``{Transverse momentum spectra and
  nuclear modification factors of charged particles in pp, p-Pb and Pb-Pb
  collisions at the LHC}'',} \textit{ JHEP} \textbf{ 11} (2018) 013,
  \href{http://dx.doi.org/10.1007/JHEP11(2018)013}{\doi{10.1007/JHEP11(2018)013}},
  \href{http://www.arXiv.org/abs/1802.09145}{\texttt{arXiv:1802.09145}}.

\bibitem{ATLAS:2022kqu}
\hrefCMSnoop {}{{ATLAS Collaboration}, ``{Charged-hadron production in $\pp$,
  $\Pp$+Pb, Pb+Pb, and Xe+Xe collisions at $\sqrtsNN=5\TeV$ with the ATLAS
  detector at the LHC}'',} \textit{ JHEP} \textbf{ 07} (2023) 074,
  \href{http://dx.doi.org/10.1007/JHEP07(2023)074}{\doi{10.1007/JHEP07(2023)074}},
  \href{http://www.arXiv.org/abs/2211.15257}{\texttt{arXiv:2211.15257}}.

\bibitem{CMS:2018yyx}
\hrefCMSnoop {}{{CMS Collaboration}, ``{Charged-particle nuclear modification
  factors in XeXe collisions at $ \sqrtsNN = 5.44\TeV$}'',} \textit{ JHEP}
  \textbf{ 10} (2018) 138,
  \href{http://dx.doi.org/10.1007/JHEP10(2018)138}{\doi{10.1007/JHEP10(2018)138}},
  \href{http://www.arXiv.org/abs/1809.00201}{\texttt{arXiv:1809.00201}}.

\bibitem{ALICE:2018hza}
\hrefCMSnoop {}{{ALICE Collaboration}, ``{Transverse momentum spectra and
  nuclear modification factors of charged particles in Xe-Xe collisions at
  $\sqrtsNN =5.44\TeV$}'',} \textit{ Phys. Lett. B} \textbf{ 788} (2019) 166,
  \href{http://dx.doi.org/10.1016/j.physletb.2018.10.052}{\doi{10.1016/j.physletb.2018.10.052}},
  \href{http://www.arXiv.org/abs/1805.04399}{\texttt{arXiv:1805.04399}}.

\bibitem{CMS:2012qk}
\hrefCMSnoop {}{{CMS Collaboration}, ``{Observation of long-range near-side
  angular correlations in proton-lead collisions at the LHC}'',} \textit{ Phys.
  Lett. B} \textbf{ 718} (2013) 795,
  \href{http://dx.doi.org/10.1016/j.physletb.2012.11.025}{\doi{10.1016/j.physletb.2012.11.025}},
  \href{http://www.arXiv.org/abs/1210.5482}{\texttt{arXiv:1210.5482}}.

\bibitem{CMS:2019isl}
\hrefCMSnoop {}{{CMS Collaboration}, ``{Strange hadron production in pp and pPb
  collisions at $\sqrtsNN= 5.02\TeV$}'',} \textit{ Phys. Rev. C} \textbf{ 101}
  (2020) 064906,
  \href{http://dx.doi.org/10.1103/PhysRevC.101.064906}{\doi{10.1103/PhysRevC.101.064906}},
  \href{http://www.arXiv.org/abs/1910.04812}{\texttt{arXiv:1910.04812}}.

\bibitem{ALICE:2016fzo}
\hrefCMSnoop {}{{ALICE Collaboration}, ``{Enhanced production of multi-strange
  hadrons in high-multiplicity proton-proton collisions}'',} \textit{ Nature
  Phys.} \textbf{ 13} (2017) 535,
  \href{http://dx.doi.org/10.1038/nphys4111}{\doi{10.1038/nphys4111}},
  \href{http://www.arXiv.org/abs/1606.07424}{\texttt{arXiv:1606.07424}}.

\bibitem{Nagle:2018nvi}
\hrefCMSnoop {}{J.~L. Nagle and W.~A. Zajc, ``{Small system collectivity in
  relativistic hadronic and nuclear collisions}'',} \textit{ Ann. Rev. Nucl.
  Part. Sci.} \textbf{ 68} (2018) 211,
  \href{http://dx.doi.org/10.1146/annurev-nucl-101916-123209}{\doi{10.1146/annurev-nucl-101916-123209}},
  \href{http://www.arXiv.org/abs/1801.03477}{\texttt{arXiv:1801.03477}}.

\bibitem{CMS:2025jbv}
\hrefCMSnoop {}{{CMS Collaboration}, ``{Search for jet quenching with dijets
  from high-multiplicity pPb collisions at $\sqrtsNN = 8.16\TeV$}'',} \textit{
  JHEP} \textbf{ 07} (2025) 118,
  \href{http://dx.doi.org/10.1007/JHEP07(2025)118}{\doi{10.1007/JHEP07(2025)118}},
  \href{http://www.arXiv.org/abs/2504.08507}{\texttt{arXiv:2504.08507}}.

\bibitem{CMS:2015ved}
\hrefCMSnoop {}{{CMS Collaboration}, ``{Nuclear effects on the transverse
  momentum spectra of charged particles in pPb collisions at $\sqrtsNN
  =5.02\TeV$}'',} \textit{ Eur. Phys. J. C} \textbf{ 75} (2015) 237,
  \href{http://dx.doi.org/10.1140/epjc/s10052-015-3435-4}{\doi{10.1140/epjc/s10052-015-3435-4}},
  \href{http://www.arXiv.org/abs/1502.05387}{\texttt{arXiv:1502.05387}}.

\bibitem{ATLAS:2022iyq}
\hrefCMSnoop {}{{ATLAS Collaboration}, ``{Strong constraints on jet quenching
  in centrality-dependent p+Pb collisions at $5.02\TeV$ from ATLAS}'',}
  \textit{ Phys. Rev. Lett.} \textbf{ 131} (2023) 072301,
  \href{http://dx.doi.org/10.1103/PhysRevLett.131.072301}{\doi{10.1103/PhysRevLett.131.072301}},
  \href{http://www.arXiv.org/abs/2206.01138}{\texttt{arXiv:2206.01138}}.

\bibitem{ALICE:2017svf}
\hrefCMSnoop {}{{ALICE Collaboration}, ``{Constraints on jet quenching in p-Pb
  collisions at $\sqrtsNN = 5.02\TeV$ measured by the event-activity dependence
  of semi-inclusive hadron-jet distributions}'',} \textit{ Phys. Lett. B}
  \textbf{ 783} (2018) 95,
  \href{http://dx.doi.org/10.1016/j.physletb.2018.05.059}{\doi{10.1016/j.physletb.2018.05.059}},
  \href{http://www.arXiv.org/abs/1712.05603}{\texttt{arXiv:1712.05603}}.

\bibitem{ALICE:2014nqx}
\hrefCMSnoop {}{{ALICE Collaboration}, ``{Transverse momentum dependence of
  inclusive primary charged-particle production in p-Pb collisions at
  $\sqrtsNN=5.02\TeV$}'',} \textit{ Eur. Phys. J. C} \textbf{ 74} (2014) 3054,
  \href{http://dx.doi.org/10.1140/epjc/s10052-014-3054-5}{\doi{10.1140/epjc/s10052-014-3054-5}},
  \href{http://www.arXiv.org/abs/1405.2737}{\texttt{arXiv:1405.2737}}.

\bibitem{PHENIX:2023dxl}
\hrefCMSnoop {}{{PHENIX} Collaboration, ``Disentangling centrality bias and
  final-state effects in the production of high-\pt neutral pions using direct
  photon in {d+Au} collisions at {$\sqrtsNN = 200\GeV$}'',} \textit{ Phys. Rev.
  Lett.} \textbf{ 134} (2025) 022302,
  \href{http://dx.doi.org/10.1103/PhysRevLett.134.022302}{\doi{10.1103/PhysRevLett.134.022302}},
  \href{http://www.arXiv.org/abs/2303.12899}{\texttt{arXiv:2303.12899}}.

\bibitem{Perepelitsa:2024eik}
\hrefCMSnoop {}{D.~V. Perepelitsa, ``{Contribution to differential \PGpz and
  {$\PGg_{\text{dir}}$} modification in small systems from color fluctuation
  effects}'',} \textit{ Phys. Rev. C} \textbf{ 110} (2024) L011901,
  \href{http://dx.doi.org/10.1103/PhysRevC.110.L011901}{\doi{10.1103/PhysRevC.110.L011901}},
  \href{http://www.arXiv.org/abs/2404.17660}{\texttt{arXiv:2404.17660}}.

\bibitem{Huss:2020dwe}
A.~Huss\hrefCMSnoop {}{ { et~al.}, ``{Discovering partonic rescattering in
  light nucleus collisions}'',} \textit{ Phys. Rev. Lett.} \textbf{ 126} (2021)
  192301,
  \href{http://dx.doi.org/10.1103/PhysRevLett.126.192301}{\doi{10.1103/PhysRevLett.126.192301}},
  \href{http://www.arXiv.org/abs/2007.13754}{\texttt{arXiv:2007.13754}}.

\bibitem{Brewer:2021kiv}
\href {https://cds.cern.ch/record/2753575}{J.~Brewer, A.~Mazeliauskas, and
  W.~van~der Schee, ``{Opportunities of OO and $p$O collisions at the LHC}'',}
  Technical Report CERN-TH-2021-028, 2021.
\newblock
  \href{http://www.arXiv.org/abs/2103.01939}{\texttt{arXiv:2103.01939}}.

\bibitem{Schenke:2020mbo}
\hrefCMSnoop {}{B.~Schenke, C.~Shen, and P.~Tribedy, ``{Running the gamut of
  high energy nuclear collisions}'',} \textit{ Phys. Rev. C} \textbf{ 102}
  (2020) 044905,
  \href{http://dx.doi.org/10.1103/PhysRevC.102.044905}{\doi{10.1103/PhysRevC.102.044905}},
  \href{http://www.arXiv.org/abs/2005.14682}{\texttt{arXiv:2005.14682}}.

\bibitem{Huss:2020whe}
A.~Huss\hrefCMSnoop {}{ { et~al.}, ``{Predicting parton energy loss in small
  collision systems}'',} \textit{ Phys. Rev. C} \textbf{ 103} (2021) 054903,
  \href{http://dx.doi.org/10.1103/PhysRevC.103.054903}{\doi{10.1103/PhysRevC.103.054903}},
  \href{http://www.arXiv.org/abs/2007.13758}{\texttt{arXiv:2007.13758}}.

\bibitem{Zakharov:2021uza}
\hrefCMSnoop {}{B.~G. Zakharov, ``{Jet quenching from heavy to light ion
  collisions}'',} \textit{ JHEP} \textbf{ 09} (2021) 087,
  \href{http://dx.doi.org/10.1007/JHEP09(2021)087}{\doi{10.1007/JHEP09(2021)087}},
  \href{http://www.arXiv.org/abs/2105.09350}{\texttt{arXiv:2105.09350}}.

\bibitem{Ke:2022gkq}
\hrefCMSnoop {}{W.~Ke and I.~Vitev, ``{Searching for QGP droplets with high-\pt
  hadrons and heavy flavor}'',} \textit{ Phys. Rev. C} \textbf{ 107} (2023)
  064903,
  \href{http://dx.doi.org/10.1103/PhysRevC.107.064903}{\doi{10.1103/PhysRevC.107.064903}},
  \href{http://www.arXiv.org/abs/2204.00634}{\texttt{arXiv:2204.00634}}.

\bibitem{dEnterria:2020dwq}
\hrefCMSnoop {}{D.~d'Enterria and C.~Loizides, ``{Progress in the Glauber model
  at collider energies}'',} \textit{ Ann. Rev. Nucl. Part. Sci.} \textbf{ 71}
  (2021) 315,
  \href{http://dx.doi.org/10.1146/annurev-nucl-102419-060007}{\doi{10.1146/annurev-nucl-102419-060007}},
  \href{http://www.arXiv.org/abs/2011.14909}{\texttt{arXiv:2011.14909}}.

\bibitem{Loizides:2016djv}
\hrefCMSnoop {}{C.~Loizides, ``{Glauber modeling of high-energy nuclear
  collisions at the subnucleon level}'',} \textit{ Phys. Rev. C} \textbf{ 94}
  (2016) 024914,
  \href{http://dx.doi.org/10.1103/PhysRevC.94.024914}{\doi{10.1103/PhysRevC.94.024914}},
  \href{http://www.arXiv.org/abs/1603.07375}{\texttt{arXiv:1603.07375}}.

\bibitem{hepdata}
\hrefCMSnoop {}{``{HEPD}ata record for this analysis'',} 2025.
\newblock
  \href{http://dx.doi.org/10.17182/hepdata.165512}{\doi{10.17182/hepdata.165512}}.

\bibitem{CMS:2008xjf}
\hrefCMSnoop {}{{CMS Collaboration}, ``The {CMS} experiment at the {CERN}
  {LHC}'',} \textit{ JINST} \textbf{ 3} (2008) S08004,
  \href{http://dx.doi.org/10.1088/1748-0221/3/08/S08004}{\doi{10.1088/1748-0221/3/08/S08004}}.

\bibitem{CMS:2023gfb}
\hrefCMSnoop {}{{CMS Collaboration}, ``{Development of the CMS detector for the
  CERN LHC Run 3}'',} \textit{ JINST} \textbf{ 19} (2024) P05064,
  \href{http://dx.doi.org/10.1088/1748-0221/19/05/P05064}{\doi{10.1088/1748-0221/19/05/P05064}},
  \href{http://www.arXiv.org/abs/2309.05466}{\texttt{arXiv:2309.05466}}.

\bibitem{vanderMeer:296752}
\href {https://cds.cern.ch/record/296752}{S.~van~der Meer, ``{Calibration of
  the effective beam height in the ISR}'',} Technical Report CERN-ISR-PO-68-31,
  ISR-PO-68-31, 1968.

\bibitem{CMS:2025rzq}
\hrefCMSnoop {}{{CMS Collaboration}, ``{Luminosity measurement for lead-lead
  collisions at $\sqrtsNN = 5.02\TeV$ in 2015 and 2018 at CMS}'',} 2025.
  \href{http://www.arXiv.org/abs/2503.03946}{\texttt{arXiv:2503.03946}}.
  Submitted to \textit{Eur. Phys. J. C..}

\bibitem{CMS:2023pad}
\hrefCMSnoop {}{{CMS Collaboration}, ``{Luminosity determination using Z boson
  production at the CMS experiment}'',} \textit{ Eur. Phys. J. C} \textbf{ 84}
  (2024) 26,
  \href{http://dx.doi.org/10.1140/epjc/s10052-023-12268-2}{\doi{10.1140/epjc/s10052-023-12268-2}},
  \href{http://www.arXiv.org/abs/2309.01008}{\texttt{arXiv:2309.01008}}.

\bibitem{CMS:2014pgm}
\hrefCMSnoop {}{{CMS Collaboration}, ``{Description and performance of track
  and primary-vertex reconstruction with the CMS tracker}'',} \textit{ JINST}
  \textbf{ 9} (2014) P10009,
  \href{http://dx.doi.org/10.1088/1748-0221/9/10/P10009}{\doi{10.1088/1748-0221/9/10/P10009}},
  \href{http://www.arXiv.org/abs/1405.6569}{\texttt{arXiv:1405.6569}}.

\bibitem{Baltz:2007kq}
\hrefCMSnoop {}{A.~J. Baltz { et~al.}, ``{The physics of ultraperipheral
  collisions at the LHC}'',} \textit{ Phys. Rept.} \textbf{ 458} (2008) 1,
  \href{http://dx.doi.org/10.1016/j.physrep.2007.12.001}{\doi{10.1016/j.physrep.2007.12.001}},
  \href{http://www.arXiv.org/abs/0706.3356}{\texttt{arXiv:0706.3356}}.

\bibitem{Pshenichnov:1997un}
I.~A. Pshenichnov\hrefCMSnoop {}{ { et~al.}, ``{Nuclear multifragmentation
  induced by electromagnetic fields of ultrarelativistic heavy ions}'',}
  \textit{ Phys. Rev. C} \textbf{ 57} (1998) 1920,
  \href{http://dx.doi.org/10.1103/PhysRevC.57.1920}{\doi{10.1103/PhysRevC.57.1920}},
  \href{http://www.arXiv.org/abs/nucl-th/9711030}{\texttt{arXiv:nucl-th/9711030}}.

\bibitem{Das:2025rqb}
\hrefCMSnoop {}{S.~J. Das and A.~Baty, ``{A data-driven method to estimate
  contamination from light ion beam transmutation at colliders}'',} 2025.
  \href{http://www.arXiv.org/abs/2509.09736}{\texttt{arXiv:2509.09736}}.
  Submitted to \textit{Phys. Rev. C}.

\bibitem{Bierlich:2022pfr}
\hrefCMSnoop {}{C.~Bierlich { et~al.}, ``{A comprehensive guide to the physics
  and usage of PYTHIA 8.3}'',} \textit{ SciPost Phys. Codeb.} \textbf{ 2022}
  (2022) 8,
  \href{http://dx.doi.org/10.21468/SciPostPhysCodeb.8}{\doi{10.21468/SciPostPhysCodeb.8}},
  \href{http://www.arXiv.org/abs/2203.11601}{\texttt{arXiv:2203.11601}}.

\bibitem{Wang:1991hta}
\hrefCMSnoop {}{X.-N. Wang and M.~Gyulassy, ``{HIJING: A Monte Carlo model for
  multiple jet production in pp, pA and AA collisions}'',} \textit{ Phys. Rev.
  D} \textbf{ 44} (1991) 3501,
  \href{http://dx.doi.org/10.1103/PhysRevD.44.3501}{\doi{10.1103/PhysRevD.44.3501}}.

\bibitem{ALICE:2020jsh}
\hrefCMSnoop {}{{ALICE Collaboration}, ``{Production of light-flavor hadrons in
  pp collisions at $\sqrt{s}~=~7$ and $\sqrt{s} = 13\TeV$}'',} \textit{ Eur.
  Phys. J. C} \textbf{ 81} (2021) 256,
  \href{http://dx.doi.org/10.1140/epjc/s10052-020-08690-5}{\doi{10.1140/epjc/s10052-020-08690-5}},
  \href{http://www.arXiv.org/abs/2005.11120}{\texttt{arXiv:2005.11120}}.

\bibitem{ALICE:2016dei}
\hrefCMSnoop {}{{ALICE Collaboration}, ``{Multiplicity dependence of charged
  pion, kaon, and (anti)proton production at large transverse momentum in
  $\Pp$-Pb collisions at $\sqrtsNN = 5.02\TeV$}'',} \textit{ Phys. Lett. B}
  \textbf{ 760} (2016) 720,
  \href{http://dx.doi.org/10.1016/j.physletb.2016.07.050}{\doi{10.1016/j.physletb.2016.07.050}},
  \href{http://www.arXiv.org/abs/1601.03658}{\texttt{arXiv:1601.03658}}.

\bibitem{ALICE:2019hno}
\hrefCMSnoop {}{{ALICE Collaboration}, ``{Production of charged pions, kaons,
  and (anti-)protons in Pb-Pb and inelastic $\Pp\Pp$ collisions at $\sqrtsNN =
  5.02\TeV$}'',} \textit{ Phys. Rev. C} \textbf{ 101} (2020) 044907,
  \href{http://dx.doi.org/10.1103/PhysRevC.101.044907}{\doi{10.1103/PhysRevC.101.044907}},
  \href{http://www.arXiv.org/abs/1910.07678}{\texttt{arXiv:1910.07678}}.

\bibitem{CMS-PAS-HIN-25-010}
\href {https://cds.cern.ch/record/2942012/files/HIN-25-010-pas.pdf}{{CMS
  Collaboration}, ``{First measurement of pseudorapidity distributions of
  charged hadrons in oxygen-oxygen collisions at $\sqrtsNN = 5.36\TeV$ with
  CMS}'',} CMS Physics Analysis Summary CMS-PAS-HIN-25-010, 2025.

\bibitem{CMS-PAS-TRK-10-002}
\href {https://cds.cern.ch/record/1279139/files/TRK-10-002-pas.pdf}{{CMS
  Collaboration}, ``Measurement of tracking efficiency'',} CMS Physics Analysis
  Summary CMS-PAS-TRK-10-002, 2010.

\bibitem{CMS:2011mry}
\hrefCMSnoop {}{{CMS Collaboration}, ``{Charged particle transverse momentum
  spectra in $\Pp\Pp$ collisions at $\sqrt{s} = 0.9$ and 7 TeV}'',} \textit{
  JHEP} \textbf{ 08} (2011) 086,
  \href{http://dx.doi.org/10.1007/JHEP08(2011)086}{\doi{10.1007/JHEP08(2011)086}},
  \href{http://www.arXiv.org/abs/1104.3547}{\texttt{arXiv:1104.3547}}.

\bibitem{CMS:2018nhd}
\hrefCMSnoop {}{{CMS Collaboration}, ``{Measurement of charged particle spectra
  in minimum-bias events from proton\textendash{}proton collisions at
  $\sqrt{s}=13\,\text {TeV} $}'',} \textit{ Eur. Phys. J. C} \textbf{ 78}
  (2018) 697,
  \href{http://dx.doi.org/10.1140/epjc/s10052-018-6144-y}{\doi{10.1140/epjc/s10052-018-6144-y}},
  \href{http://www.arXiv.org/abs/1806.11245}{\texttt{arXiv:1806.11245}}.

\bibitem{Antreasyan:1978cw}
D.~Antreasyan\hrefCMSnoop {}{ { et~al.}, ``{Production of hadrons at large
  transverse momentum in 200, 300 and 400 GeV pp and pn collisions}'',}
  \textit{ Phys. Rev. D} \textbf{ 19} (1979) 764,
  \href{http://dx.doi.org/10.1103/PhysRevD.19.764}{\doi{10.1103/PhysRevD.19.764}}.

\bibitem{ATLAS:2025ztg}
\hrefCMSnoop {}{{ATLAS Collaboration}, ``{Evidence for the collective nature of
  radial flow in Pb+Pb collisions with the ATLAS detector}'',} 2025.
  \href{http://www.arXiv.org/abs/2503.24125}{\texttt{arXiv:2503.24125}}.
  Submitted to PRL.

\bibitem{ALICE:2025iud}
\hrefCMSnoop {}{{ALICE Collaboration}, ``{Long-range transverse momentum
  correlations and radial flow in Pb$-$Pb collisions at the LHC}'',} 2025.
  \href{http://www.arXiv.org/abs/2504.04796}{\texttt{arXiv:2504.04796}}.
  Submitted to PRL.

\bibitem{Xie:2022ght}
\hrefCMSnoop {}{M.~Xie, W.~Ke, H.~Zhang, and X.-N. Wang,
  ``{Information-field-based global Bayesian inference of the jet transport
  coefficient}'',} \textit{ Phys. Rev. C} \textbf{ 108} (2023) L011901,
  \href{http://dx.doi.org/10.1103/PhysRevC.108.L011901}{\doi{10.1103/PhysRevC.108.L011901}},
  \href{http://www.arXiv.org/abs/2206.01340}{\texttt{arXiv:2206.01340}}.

\bibitem{Shi:2018izg}
\hrefCMSnoop {}{S.~Shi, J.~Liao, and M.~Gyulassy, ``{Global constraints from
  RHIC and LHC on transport properties of QCD fluids in CUJET/CIBJET
  framework}'',} \textit{ Chin. Phys. C} \textbf{ 43} (2019) 044101,
  \href{http://dx.doi.org/10.1088/1674-1137/43/4/044101}{\doi{10.1088/1674-1137/43/4/044101}},
  \href{http://www.arXiv.org/abs/1808.05461}{\texttt{arXiv:1808.05461}}.

\bibitem{Shi:2018lsf}
\hrefCMSnoop {}{S.~Shi, J.~Liao, and M.~Gyulassy, ``{Probing the Color
  Structure of the Perfect QCD Fluids via Soft-Hard-Event-by-Event Azimuthal
  Correlations}'',} \textit{ Chin. Phys. C} \textbf{ 42} (2018) 104104,
  \href{http://dx.doi.org/10.1088/1674-1137/42/10/104104}{\doi{10.1088/1674-1137/42/10/104104}},
  \href{http://www.arXiv.org/abs/1804.01915}{\texttt{arXiv:1804.01915}}.

\bibitem{Casalderrey-Solana:2014bpa}
J.~Casalderrey-Solana\hrefCMSnoop {}{ { et~al.}, ``{A hybrid strong/weak
  coupling approach to jet quenching}'',} \textit{ JHEP} \textbf{ 10} (2014)
  019,
  \href{http://dx.doi.org/10.1007/JHEP10(2014)019}{\doi{10.1007/JHEP10(2014)019}},
  \href{http://www.arXiv.org/abs/1405.3864}{\texttt{arXiv:1405.3864}}.
  [Erratum: \textit{JHEP} \textbf{09} (2015) 175,
  \url{doi:10.1007/JHEP09(2015)175}].

\bibitem{Gebhard:2024flv}
\hrefCMSnoop {}{J.~Gebhard, A.~Mazeliauskas, and A.~Takacs, ``{No-quenching
  baseline for energy loss signals in oxygen-oxygen collisions}'',} \textit{
  JHEP} \textbf{ 04} (2025) 034,
  \href{http://dx.doi.org/10.1007/JHEP04(2025)034}{\doi{10.1007/JHEP04(2025)034}},
  \href{http://www.arXiv.org/abs/2410.22405}{\texttt{arXiv:2410.22405}}.

\bibitem{Pablos:2025cli}
\hrefCMSnoop {}{D.~Pablos and A.~Takacs, ``{Bayesian constraints on
  pre-equilibrium jet quenching and predictions for oxygen collisions}'',}
  2025. \href{http://www.arXiv.org/abs/2509.19430}{\texttt{arXiv:2509.19430}}.

\bibitem{vanderSchee:2025hoe}
W.~van~der Schee\hrefCMSnoop {}{ { et~al.}, ``{Three models for charged hadron
  nuclear modification from light to heavy ions}'',} 2025.
  \href{http://www.arXiv.org/abs/2509.04299}{\texttt{arXiv:2509.04299}}.

\bibitem{Faraday:2025pto}
\hrefCMSnoop {}{C.~Faraday and W.~A. Horowitz, ``{Statistical analysis of pQCD
  energy loss across system size, flavor, $\sqrtsNN$, and $\pt$}'',} \textit{
  JHEP} \textbf{ 11} (2025) 019,
  \href{http://dx.doi.org/10.1007/JHEP11(2025)019}{\doi{10.1007/JHEP11(2025)019}},
  \href{http://www.arXiv.org/abs/2505.14568}{\texttt{arXiv:2505.14568}}.

\bibitem{Mazeliauskas:2025clt}
\hrefCMSnoop {}{A.~Mazeliauskas, ``{Energy loss baseline for light hadrons in
  oxygen-oxygen collisions at $\sqrtsNN = 5.36\TeV$}'',} 2025.
  \href{http://www.arXiv.org/abs/2509.07008}{\texttt{arXiv:2509.07008}}.

\bibitem{Eskola:2021nhw}
\hrefCMSnoop {}{K.~J. Eskola, P.~Paakkinen, H.~Paukkunen, and C.~A. Salgado,
  ``{EPPS21: a global QCD analysis of nuclear PDFs}'',} \textit{ Eur. Phys. J.
  C} \textbf{ 82} (2022) 413,
  \href{http://dx.doi.org/10.1140/epjc/s10052-022-10359-0}{\doi{10.1140/epjc/s10052-022-10359-0}},
  \href{http://www.arXiv.org/abs/2112.12462}{\texttt{arXiv:2112.12462}}.

\bibitem{suppMatBib}
\hrefCMSnoop {}{``Supplemental material: Comparison to different nPDF
  models''.}
\newblock [URL will be inserted by publisher at publication].

\bibitem{Eskola:2009uj}
\hrefCMSnoop {}{K.~J. Eskola, H.~Paukkunen, and C.~A. Salgado, ``{EPS09: A new
  generation of NLO and LO nuclear parton distribution functions}'',} \textit{
  JHEP} \textbf{ 04} (2009) 065,
  \href{http://dx.doi.org/10.1088/1126-6708/2009/04/065}{\doi{10.1088/1126-6708/2009/04/065}},
  \href{http://www.arXiv.org/abs/0902.4154}{\texttt{arXiv:0902.4154}}.

\bibitem{Eskola:2016oht}
\hrefCMSnoop {}{K.~J. Eskola, P.~Paakkinen, H.~Paukkunen, and C.~A. Salgado,
  ``{EPPS16: Nuclear parton distributions with LHC data}'',} \textit{ Eur.
  Phys. J. C} \textbf{ 77} (2017) 163,
  \href{http://dx.doi.org/10.1140/epjc/s10052-017-4725-9}{\doi{10.1140/epjc/s10052-017-4725-9}},
  \href{http://www.arXiv.org/abs/1612.05741}{\texttt{arXiv:1612.05741}}.

\bibitem{Pang:2012he}
\hrefCMSnoop {}{L.~Pang, Q.~Wang, and X.-N. Wang, ``{Effects of initial flow
  velocity fluctuation in event-by-event (3+1)D hydrodynamics}'',} \textit{
  Phys. Rev. C} \textbf{ 86} (2012) 024911,
  \href{http://dx.doi.org/10.1103/PhysRevC.86.024911}{\doi{10.1103/PhysRevC.86.024911}},
  \href{http://www.arXiv.org/abs/1205.5019}{\texttt{arXiv:1205.5019}}.

\bibitem{Djordjevic:2003zk}
\hrefCMSnoop {}{M.~Djordjevic and M.~Gyulassy, ``{Heavy quark radiative energy
  loss in QCD matter}'',} \textit{ Nucl. Phys. A} \textbf{ 733} (2004) 265,
  \href{http://dx.doi.org/10.1016/j.nuclphysa.2003.12.020}{\doi{10.1016/j.nuclphysa.2003.12.020}},
  \href{http://www.arXiv.org/abs/nucl-th/0310076}{\texttt{arXiv:nucl-th/0310076}}.

\bibitem{Gyulassy:1999zd}
\hrefCMSnoop {}{M.~Gyulassy, P.~Levai, and I.~Vitev, ``{Jet quenching in thin
  quark gluon plasmas. 1. Formalism}'',} \textit{ Nucl. Phys. B} \textbf{ 571}
  (2000) 197,
  \href{http://dx.doi.org/10.1016/S0550-3213(99)00713-0}{\doi{10.1016/S0550-3213(99)00713-0}},
  \href{http://www.arXiv.org/abs/hep-ph/9907461}{\texttt{arXiv:hep-ph/9907461}}.

\bibitem{Mehtar-Tani:2021fud}
\hrefCMSnoop {}{Y.~Mehtar-Tani, D.~Pablos, and K.~Tywoniuk, ``{Cone-size
  dependence of jet suppression in heavy-ion collisions}'',} \textit{ Phys.
  Rev. Lett.} \textbf{ 127} (2021) 252301,
  \href{http://dx.doi.org/10.1103/PhysRevLett.127.252301}{\doi{10.1103/PhysRevLett.127.252301}},
  \href{http://www.arXiv.org/abs/2101.01742}{\texttt{arXiv:2101.01742}}.

\bibitem{Takacs:2021bpv}
\hrefCMSnoop {}{A.~Takacs and K.~Tywoniuk, ``{Quenching effects in the
  cumulative jet spectrum}'',} \textit{ JHEP} \textbf{ 10} (2021) 038,
  \href{http://dx.doi.org/10.1007/JHEP10(2021)038}{\doi{10.1007/JHEP10(2021)038}},
  \href{http://www.arXiv.org/abs/2103.14676}{\texttt{arXiv:2103.14676}}.

\bibitem{Mehtar-Tani:2024jtd}
\hrefCMSnoop {}{Y.~Mehtar-Tani, D.~Pablos, and K.~Tywoniuk, ``{Jet suppression
  and azimuthal anisotropy from RHIC to LHC}'',} \textit{ Phys. Rev. D}
  \textbf{ 110} (2024) 014009,
  \href{http://dx.doi.org/10.1103/PhysRevD.110.014009}{\doi{10.1103/PhysRevD.110.014009}},
  \href{http://www.arXiv.org/abs/2402.07869}{\texttt{arXiv:2402.07869}}.

\bibitem{Mantysaari:2025tcg}
\hrefCMSnoop {}{H.~M{\"a}ntysaari, B.~Schenke, C.~Shen, and W.~Zhao,
  ``{Collision-energy dependence in heavy-ion collisions from nonlinear QCD
  evolution}'',} \textit{ Phys. Rev. Lett.} \textbf{ 135} (2025) 022302,
  \href{http://dx.doi.org/10.1103/gf4y-p5j7}{\doi{10.1103/gf4y-p5j7}},
  \href{http://www.arXiv.org/abs/2502.05138}{\texttt{arXiv:2502.05138}}.

\bibitem{Zapp:2013vla}
\hrefCMSnoop {}{K.~C. Zapp, ``{JEWEL 2.0.0: directions for use}'',} \textit{
  Eur. Phys. J. C} \textbf{ 74} (2014) 2762,
  \href{http://dx.doi.org/10.1140/epjc/s10052-014-2762-1}{\doi{10.1140/epjc/s10052-014-2762-1}},
  \href{http://www.arXiv.org/abs/1311.0048}{\texttt{arXiv:1311.0048}}.

\bibitem{Kolbe:2023rsq}
\hrefCMSnoop {}{I.~Kolb{\'e}, ``{JEWEL on a (2+1)D background with applications
  to small systems and substructure}'',} 2023.
  \href{http://www.arXiv.org/abs/2303.14166}{\texttt{arXiv:2303.14166}}.

\bibitem{Nijs:2020ors}
\hrefCMSnoop {}{G.~Nijs, W.~van~der Schee, U.~G{\"u}rsoy, and R.~Snellings,
  ``{Transverse momentum differential global analysis of heavy-ion
  collisions}'',} \textit{ Phys. Rev. Lett.} \textbf{ 126} (2021) 202301,
  \href{http://dx.doi.org/10.1103/PhysRevLett.126.202301}{\doi{10.1103/PhysRevLett.126.202301}},
  \href{http://www.arXiv.org/abs/2010.15130}{\texttt{arXiv:2010.15130}}.

\bibitem{Nijs:2020roc}
\hrefCMSnoop {}{G.~Nijs, W.~van~der Schee, U.~G{\"u}rsoy, and R.~Snellings,
  ``{Bayesian analysis of heavy ion collisions with the heavy ion computational
  framework Trajectum}'',} \textit{ Phys. Rev. C} \textbf{ 103} (2021) 054909,
  \href{http://dx.doi.org/10.1103/PhysRevC.103.054909}{\doi{10.1103/PhysRevC.103.054909}},
  \href{http://www.arXiv.org/abs/2010.15134}{\texttt{arXiv:2010.15134}}.

\bibitem{vanderSchee:2023uii}
\hrefCMSnoop {}{W.~van~der Schee, Y.-J. Lee, G.~Nijs, and Y.~Chen, ``{Hard
  probes in isobar collisions as a probe of the neutron skin}'',} \textit{
  Phys. Lett. B} \textbf{ 856} (2024) 138953,
  \href{http://dx.doi.org/10.1016/j.physletb.2024.138953}{\doi{10.1016/j.physletb.2024.138953}},
  \href{http://www.arXiv.org/abs/2307.11836}{\texttt{arXiv:2307.11836}}.

\bibitem{CMS:2019csb}
\hrefCMSnoop {}{{CMS Collaboration}, ``{Extraction and validation of a new set
  of CMS PYTHIA8 tunes from underlying-event measurements}'',} \textit{ Eur.
  Phys. J. C} \textbf{ 80} (2020) 4,
  \href{http://dx.doi.org/10.1140/epjc/s10052-019-7499-4}{\doi{10.1140/epjc/s10052-019-7499-4}},
  \href{http://www.arXiv.org/abs/1903.12179}{\texttt{arXiv:1903.12179}}.

\bibitem{Klein:2016yzr}
S.~R. Klein\hrefCMSnoop {}{ { et~al.}, ``{STARlight: A Monte Carlo simulation
  program for ultra-peripheral collisions of relativistic ions}'',} \textit{
  Comput. Phys. Commun.} \textbf{ 212} (2017) 258,
  \href{http://dx.doi.org/10.1016/j.cpc.2016.10.016}{\doi{10.1016/j.cpc.2016.10.016}},
  \href{http://www.arXiv.org/abs/1607.03838}{\texttt{arXiv:1607.03838}}.

\bibitem{Agostinelli:2002hh}
\hrefCMSnoop {}{{GEANT4} Collaboration, ``{\GEANTfour}---a simulation
  toolkit'',} \textit{ Nucl. Instrum. Meth. A} \textbf{ 506} (2003) 250,
\href{http://dx.doi.org/10.1016/S0168-9002(03)01368-8}{\doi{10.1016/S0168-9002(03)01368-8}}.

\bibitem{AbdulKhalek:2022fyi}
R.~Abdul~Khalek\hrefCMSnoop {}{ { et~al.}, ``{nNNPDF3.0: evidence for a
  modified partonic structure in heavy nuclei}'',} \textit{ Eur. Phys. J. C}
  \textbf{ 82} (2022) 507,
  \href{http://dx.doi.org/10.1140/epjc/s10052-022-10417-7}{\doi{10.1140/epjc/s10052-022-10417-7}},
  \href{http://www.arXiv.org/abs/2201.12363}{\texttt{arXiv:2201.12363}}.

\bibitem{Helenius:2021tof}
\hrefCMSnoop {}{I.~Helenius, M.~Walt, and W.~Vogelsang, ``{NNLO nuclear parton
  distribution functions with electroweak-boson production data from the
  LHC}'',} \textit{ Phys. Rev. D} \textbf{ 105} (2022) 094031,
  \href{http://dx.doi.org/10.1103/PhysRevD.105.094031}{\doi{10.1103/PhysRevD.105.094031}},
  \href{http://www.arXiv.org/abs/2112.11904}{\texttt{arXiv:2112.11904}}.

\bibitem{Klasen:2023uqj}
\hrefCMSnoop {}{M.~Klasen and H.~Paukkunen, ``{Nuclear PDFs after the first
  decade of LHC data}'',} \textit{ Ann. Rev. Nucl. Part. Sci.} \textbf{ 74}
  (2024) 49,
  \href{http://dx.doi.org/10.1146/annurev-nucl-102122-022747}{\doi{10.1146/annurev-nucl-102122-022747}},
  \href{http://www.arXiv.org/abs/2311.00450}{\texttt{arXiv:2311.00450}}.

\bibitem{Paakkinen:2021jjp}
\hrefCMSnoop {}{P.~Paakkinen, ``{Light-nuclei gluons from dijet production in
  proton-oxygen collisions}'',} \textit{ Phys. Rev. D} \textbf{ 105} (2022)
  L031504,
  \href{http://dx.doi.org/10.1103/PhysRevD.105.L031504}{\doi{10.1103/PhysRevD.105.L031504}},
  \href{http://www.arXiv.org/abs/2111.05368}{\texttt{arXiv:2111.05368}}.

\end{thebibliography}\endgroup
